\documentclass[acmsmall,10pt,screen]{acmart}

\startPage{1}

\usepackage{booktabs}   
\usepackage{subcaption} 

\usepackage{stmaryrd}
\usepackage{amsmath}
\usepackage{amsfonts}
\usepackage{epsfig}
\usepackage{graphicx}
\usepackage{algorithmic}
\usepackage{algorithm}
\usepackage{xspace}
\usepackage{listings}
\usepackage{array}
\usepackage{multirow}
\usepackage{wrapfig}
\usepackage{color}
\usepackage{xcolor}
\usepackage{tikz}
\usepackage{program}
\usepackage{adjustbox}
\usepackage{longtable}

\begin{document}

\newcommand{\note}[1]{{\color{blue}{\small\sf [#1]}}}
\newcommand{\com}[1]{{\color{red}{\small\sf [#1]}}}
\newcommand{\sys}{\textsc{WebRelate}}
\definecolor{mygray}{gray}{0.7}
\newcommand{\h}[1]{\colorbox{mygray}{#1}}
\renewcommand\t[1]{\texttt{#1}}

\newcommand{\lweb}{$L_w$}
\newcommand\iftr[1]{}
\newcommand{\seclabel}[1]{\label{sec:#1}}
\newcommand{\exlabel}[1]{\label{ex:#1}}
\newcommand{\tablabel}[1]{\label{tab:#1}}
\newcommand{\tabref}[1]{Table~\ref{tab:#1}}
\newcommand{\exref}[1]{Ex~\ref{ex:#1}}
\newcommand{\figlabel}[1]{\label{fig:#1}}
\newcommand{\longfigref}[1]{Fig.~\ref{fig:#1}}
\newcommand{\longsecref}[1]{\S~\ref{sec:#1}}
\newcommand{\shortfigref}[1]{Fig.~\ref{fig:#1}}
\newcommand{\eqqlabel}[1]{\label{eq:#1}}
\newcommand{\shorteqqref}[1]{\eqref{eq:#1}}
\newcommand{\mediumeqqref}[1]{Eq.~\eqref{eq:#1}}
\newcommand{\longeqqref}[1]{Equation~\eqref{eq:#1}}
\newcommand{\secref}[1]{\longsecref{#1}}
\newcommand{\figref}[1]{\longfigref{#1}}
\newcommand{\eqqref}[1]{\longeqqref{#1}}
\newcommand{\linelabel}[1]{\label{line:#1}}
\newcommand{\lineref}[1]{Line~\ref{line:#1}}
\newcommand{\linesref}[2]{Lines~\ref{line:#1}--\ref{line:#2}}

\newcommand{\len}{\t{len}}

\newcommand{\HierarchicalSearch}{\t{HierarchicalSearch}}
\newcommand{\entityi}{i}
\newcommand{\entityo}{o}
\newcommand{\Dag}{\t{Dag}}
\newcommand{\GenProg}{\t{GenProg}}
\newcommand{\isSubstr}{\t{isSubstr}}
\newcommand{\isConst}{\t{isConst}}
\newcommand{\delim}{\t{D}}
\newcommand{\GenDag}{\t{GenDag}}
\newcommand{\Intersect}{\t{Intersect}}
\newcommand{\nodes}{\mathcal{V}}
\newcommand{\splitIdx}{P}
\newcommand{\splitC}{\t{split}}
\newcommand{\edges}{\mathcal{E}}
\newcommand{\edgemap}{w}
\newcommand{\set}{\t{set}}
\newcommand{\add}{\t{add}}
\newcommand{\GenSubstr}{\t{GenSubstr}}
\newcommand{\GenConst}{\t{GenConst}}
\newcommand{\nodestart}{\nu_s}
\newcommand{\nodetarget}{\nu_t}
\newcommand{\nullC}{\t{null}}
\newcommand{\mapC}{\t{map}}

\newcommand{\onlyWords}{\t{onlyWords}}
\newcommand{\andC}{\t{and}}
\newcommand{\true}{\t{true}}
\newcommand{\false}{\t{false}}
\newcommand{\substrInWords}{\t{substrInWords}}

\newcommand{\entityurls}{P}
\newcommand{\entityurl}{p}
\newcommand{\entityp}{f}
\newcommand{\run}{\t{run}}
\newcommand{\matches}{\t{matches}}
\newcommand{\Program}{\t{Program}}
\newcommand{\TopRank}{\t{TopRank}}
\newcommand{\entityplist}{F}
\newcommand{\runList}{\t{run}}
\newcommand{\entityr}{r}
\newcommand{\breakC}{\t{break}}
\newcommand{\RankD}{\t{RankD}}

\newcommand{\entitye}{e}
\newcommand{\Edges}{\t{Edges}}
\newcommand{\Contains}{\t{Contains}}
\newcommand{\Rank}{\t{Rank}}
\newcommand{\LearnDags}{\t{LearnDags}}
\newcommand{\LearnDag}{\t{LearnDag}}
\newcommand{\entityweights}{W}
\newcommand{\LearnFilters}{\t{LearnFilters}}
\newcommand{\LearnFilter}{\t{LearnFilter}}
\newcommand{\entitydags}{\t{dags}}

\newcommand{\dom}{w}
\newcommand{\semw}[1]{\sem{#1}(\dom)}
\newcommand{\Map}{\t{Map}}
\newcommand{\Flatten}{\t{Flatten}}
\newcommand{\Filter}{\t{Filter}}
\newcommand{\entitynodesem}{\gamma}
\newcommand{\semn}[1]{\sem{#1}(\entitynodesem)}
\newcommand{\semnl}[1]{\sem{#1}(\{\entitynodesem\})}
\newcommand{\nodelist}{\{\entitynodesem_k\}_k}
\newcommand{\semnll}[1]{\sem{#1}(\nodelist)}

\newcommand{\LearnTree}{\t{LearnTree}}
\newcommand{\ReplaceStrWithDag}{\t{ReplaceStrWithDag}}
\newcommand{\entitynodes}{\t{nodes}}
\newcommand{\entitynode}{\t{n}}
\newcommand{\entitychildren}{C}
\newcommand{\IntersectNode}{\t{IntersectNode}}
\newcommand{\IntersectChildren}{\t{IntersectChildren}}
\newcommand{\entitynodea}{\t{node}_1}
\newcommand{\entitynodeb}{\t{node}_2}

\newcommand{\entitypredicates}{P}
\newcommand{\emptysetC}{\phi}
\newcommand{\notC}{\t{not}}
\newcommand{\orC}{\t{or}}
\newcommand{\Attr}{\t{Attr}}
\newcommand{\entityattr}{\t{attr}}
\newcommand{\Attributes}{\t{Attributes}}
\newcommand{\intersectAttribute}{\t{IntersectAttribute}}
\newcommand{\Node}{\t{Node}}
\newcommand{\Name}{\t{Name}}

\newcommand{\numInputs}{N}
\newcommand{\entitystr}{\t{str}}
\newcommand{\AllStrings}{\t{AllStrings}}
\newcommand{\entitydag}{d}
\newcommand{\Replace}{\t{Replace}}

\newcommand{\isLeftMost}{\t{isLeftMost}}
\newcommand{\isRightMost}{\t{isRightMost}}
\newcommand{\isLeaf}{\t{isLeaf}}
\newcommand{\CountPred}{\t{Count}}
\newcommand{\Left}{\t{"Left"}}
\newcommand{\Right}{\t{"Right"}}
\newcommand{\Child}{\t{"Child"}}
\newcommand{\Ancestor}{\t{"Ancestor"}}

\newcommand{\Children}{\t{Children}}
\newcommand{\entitychild}{c}
\newcommand{\Axis}{\t{Axis}}

\newcommand{\entityvalue}{\t{value}}
\newcommand{\AttrValue}{\t{AttrValue}}

\newcommand{\FindMinConstraints}{\t{FindMinPredicates}}

\newcommand{\entitydist}{d}
\newcommand{\distance}{\t{distance}}
\newcommand{\entityconstraints}{\gamma}
\newcommand{\entitypage}{w}
\newcommand{\entityhtmlnode}{o}
\newcommand{\entitytestpage}{t}
\newcommand{\satisfiesEx}{\t{satisfiesEx}}
\newcommand{\satisfiesTest}{\t{satisfiesTest}}
\newcommand{\pairs}{\t{pairs}}
\newcommand{\entitypath}{\t{path}}
\newcommand{\entitypred}{\pi}
\newcommand{\RankPredicates}{\t{RankP}}
\newcommand{\pair}{\t{pair}}
\newcommand{\remove}{\t{remove}}
\newcommand{\Predicates}{\t{Predicates}}
\newcommand{\Position}{\t{Position}}
\newcommand{\maxC}{\t{max}}
\newcommand{\equalP}{\t{"=="}}
\newcommand{\lteP}{\t{"<="}}

\newcommand{\Synth}{\t{Synth}}
\newcommand{\entityallpreds}{\Pi}
\newcommand{\LearnPredicates}{\t{LearnPredicates}}
\newcommand{\SortedIterator}{\t{SortedIterator}}
\newcommand{\UpdatePred}{\t{RefinePath}}
\newcommand{\AddPred}{\t{AddPred}}
\newcommand{\RemovePred}{\t{RemovePred}}
\newcommand{\entityilist}{I}
\newcommand{\entityhlist}{\Gamma}
\newcommand{\entityhnode}{\gamma}
\newcommand{\entitypagelist}{W}
\newcommand{\entitytest}{T}
\newcommand{\entityii}{i}
\newcommand{\RefinePredicates}{\t{RefinePredicates}}
\newcommand{\Search}{\t{SearchBestProg}}
\newcommand{\Refine}{\t{Refine}}
\newcommand{\IntersectPath}{\t{IntersectPath}}

\newcommand{\entityTAnchor}{T}
\newcommand{\entityAnchors}{A}
\newcommand{\entityAnchor}{a}
\newcommand{\entityHops}{H}
\newcommand{\TargetAnchor}{\t{TargetAnchor}}
\newcommand{\LearnAttrPred}{\t{LearnAttrPred}}
\newcommand{\LearnCountPred}{\t{LearnCountPred}}
\newcommand{\LearnNode}{\t{LearnNode}}
\newcommand{\LearnChildren}{\t{LearnChildren}}
\newcommand{\Anchor}{\t{Anchor}}
\newcommand{\LearnSiblings}{\t{LearnSiblings}}
\newcommand{\LearnAncestors}{\t{LearnAncestors}}
\newcommand{\entityhchild}{\gamma_c}
\newcommand{\LearnAnchor}{\t{LearnAnchor}}
\newcommand{\entityhop}{h}
\newcommand{\Hop}{\t{Hop}}
\newcommand{\entitydir}{\t{dir}}

\newcommand{\entitypreds}{\Pi_I}
\newcommand{\entitypathslist}{\mathcal{P}}
\newcommand{\entitypathp}{p}

\newcommand{\sem}[1]{\llbracket #1 \rrbracket_{i}}
\newcommand{\semg}[1]{\llbracket #1 \rrbracket_{i, O_u}}
\newcommand{\sems}[1]{\llbracket #1 \rrbracket_{i}}
\newcommand\semt[2]{\llbracket #1 \rrbracket_{#2}\xspace}
\newcommand{\strvar}{s}
\newcommand{\entity}{e}

\newcommand\myparagraph[1]{\textbf{#1}:}

\newcommand{\quotes}[1]{``#1''}

\newcommand{\substrlambda}{\lambda_s}
\newcommand{\constlambda}{\lambda_c}
\newcommand{\randomlambda}{\lambda_a}

\newcommand{\lurl}{L_u}

\title{WebRelate: Integrating Web Data with Spreadsheets using Examples}


\author{Jeevana Priya Inala}
\affiliation{
  \department{CSAIL}              
  \institution{MIT}            
  \streetaddress{32 Vassar Street}
  \city{Cambridge}
  \state{MA}
  \postcode{02139}
  \country{USA}
}
\email{jinala@csail.mit.edu}          

\author{Rishabh Singh}
\affiliation{
  \institution{Microsoft Research}           
  \city{Redmond}
  \state{WA}
  \country{USA}
}
\email{risin@microsoft.com}         


\begin{abstract}
Data integration between web sources and relational data is a key challenge faced by data scientists and spreadsheet users. There are two main challenges in programmatically joining web data with relational data. First, most websites do not expose a direct interface to obtain tabular data, so the user needs to formulate a logic to get to different webpages for each input row in the relational table. Second, after reaching the desired webpage, the user needs to write complex scripts to extract the relevant data, which is often conditioned on the input data. Since many data scientists and end-users come from diverse backgrounds, writing such complex regular-expression based logical scripts to perform data integration tasks is unfortunately often beyond their programming expertise.

We present $\sys$, a system that allows users to join semi-structured web data with relational data in spreadsheets using input-output examples.  $\sys$ decomposes the web data integration task into two sub-tasks of i) URL learning and ii) input-dependent web extraction. We introduce a novel synthesis paradigm called \quotes{Output-constrained Programming By Examples}, which allows us to use the finite set of possible outputs for the new inputs to efficiently constrain the search in the synthesis algorithm. We instantiate this paradigm for the two sub-tasks in $\sys$. The first sub-task generates the URLs for the webpages containing the desired data for all rows in the relational table. $\sys$ achieves this by learning a string transformation program using a few example URLs. The second sub-task uses examples of desired data to be extracted from the corresponding webpages and learns a program to extract the data for the other rows.
We design expressive domain-specific languages for URL generation and web data extraction, and  present efficient synthesis algorithms for learning programs in these DSLs from few input-output examples. We evaluate $\sys$ on 88 real-world web data integration tasks taken from online help forums and Excel product team, and show that \sys{} can learn the desired programs within few seconds using only 1 example for the majority of the tasks.
\end{abstract}

\setcopyright{acmlicensed}
\acmPrice{}
\acmDOI{10.1145/3158090}
\acmYear{2018}
\copyrightyear{2018}
\acmJournal{PACMPL}
\acmVolume{2}
\acmNumber{POPL}
\acmArticle{2}
\acmMonth{1}




\maketitle

\section{Introduction}
Data integration is a key challenge faced by many data scientists and spreadsheet end-users. Despite several recent advances in techniques for making it easier for users to perform data analysis~\cite{wrangling}, the inferences obtained from the
analyses is only as good as the information diversity of the
data. Therefore, more and more users are enriching their internal data in spreadsheets with the rich data available on the web. However,  the web data sources (websites) are typically semi-structured and in a format different from the original spreadsheet data format, and hence, performing such integration tasks requires writing complex regular-expression based data transformation and web scraping scripts. Unfortunately, a large fraction of end-users come from diverse backgrounds and writing such scripts is beyond their programming expertise~\cite{forrestor}. 
Even for experienced programmers and data scientists, writing such data integration scripts can be difficult and time consuming.


\begin{example}\exlabel{cur}
	To make the problem concrete, consider the integration task shown in \figref{ex-cur}. A user had a spreadsheet consisting of pairs of currencies and the dates of transaction (shown in \figref{ex-cur}(a)), and wanted to get the exchange rates from the web. Since there were thousands of rows in the spreadsheet, manually performing this task was prohibitively expensive for the user.
\end{example}

\begin{figure}[!htpb]
	\begin{tabular}{c c}
		\begin{minipage}{0.6\linewidth}
			\small
			\begin{center}
				\begin{tabular}{|c|c|c|c|c|}
					\hline
					& Cur 1 & Cur 2 & Date & Exchange Rate \\\hline\hline
					1 & EUR & USD & 03, November, 16 &  \\ \hline
					2 & USD & INR & 01, November, 16 & \\ \hline 
					3 & AUD & CAD & 07, October, 16 & \\ \hline
				\end{tabular}
			\end{center}
			
		\end{minipage}
		
		&
		\begin{minipage}{0.35\linewidth}
			\includegraphics[scale=0.5]{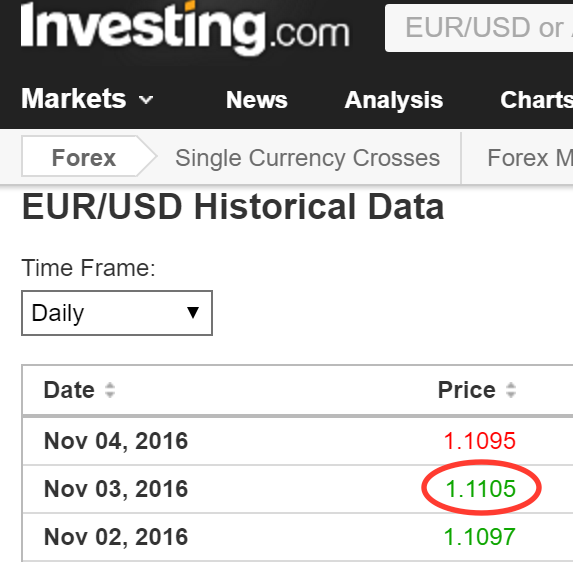}
		\end{minipage}
		\\
		(a) & (b)
	\end{tabular}
	\caption{(a) A spreadsheet containing pairs of currency symbols and dates of transaction. (b) The webpage for EUR to USD exchange rate for different dates.}
	\figlabel{ex-cur}
\end{figure} 


Previous works have explored two different strategies for automating such data integration tasks. In DataXFormer~\cite{dataxformer}, a user provides a few end-to-end input-output examples (such as $row1 \rightarrow 1.1105$ in the above example),  
 and the system uses a fully automatic approach by searching through a huge database of web forms and web tables to find a transform that is consistent with the examples. However, many websites do not expose web forms and do not have the data in a single web table. On the other end are programming by demonstration (PBD) systems such as WebCombine~\cite{webcombine} and Vegemite~\cite{vegemite} that rely on users to demonstrate how to navigate and retrieve  the desired data for a few example rows.  Although the PBD systems can handle a broader range of webpages compared to DataXFormer, they tend to put an additional burden on users to perform exact demonstrations to get to the webpage, which has been shown to be problematic for users~\cite{pbdfail}. 

 In this paper, we present an intermediate approach where a user provides a few examples of URLs of webpages that contain the data as well as examples of the desired data from these webpages, and the system automatically learns how to perform the integration task for the other rows in the spreadsheet. For instance, in the above task, the example URL for the first row is \url{http://www.investing.com/currencies/eur-usd-historical-data} and the corresponding webpage is shown in \figref{ex-cur}(b) where the user highlights the desired data. 
There are three main challenges for a system to learn a program to automate this integration task.

First, the system needs to learn a program to generate the desired URLs for the other input rows. In the above example, the intended program for learning URLs needs to select the appropriate columns from the spreadsheet, perform casing transformations, and then concatenate them with appropriate constant strings. Moreover, many URL generation programs require using string transformations on input data based on complex regular expressions.


The second challenge is to learn an extraction logic to retrieve the relevant data, which depends on the underlying DOM structure of the webpage. Additionally, for many integration scenarios, the data that needs to be extracted from the web is conditioned on the data in the table. For instance, in the above example, the currency exchange rate from the web page should be extracted based on the \t{Date} column in the table. 
There are efficient systems in the literature for wrapper induction~\cite{kushmerick97,anton05,dalvi09,stalker} that can learn robust and generalizable extraction programs from a few labeled examples, but, to the best of our knowledge, there is no previous work that can learn data extraction programs that are conditioned on the input.

The third challenge is that the common data key to join the two sources (spreadsheet and web data) might be in different formats, which requires writing additional logic to transform the data before performing the join. For instance, in the example above,  the format of the date in the web page \quotes{\t{Nov 03, 2016}} is different from the format in the spreadsheet \quotes{\t{03, November, 16}}. 

To address  the above challenges, we present $\sys$, a system that uses input-output examples to learn a program to perform the web data integration task.
The high-level overview of the system is shown in \figref{overview}. It breaks down the integration task into two sub-tasks. The first sub-task  uses the \emph{URL synthesizier} module to learn a program to generate the URLs of the webpages that contain the desired data from few example URLs. The second sub-task uses the \emph{Data synthesizer} module to learn a data extraction program to retrieve the relevant data from these URLs given a set of example data selections on the webpages.


\begin{figure}[t]
	\centering
	\includegraphics[width=0.6\textwidth]{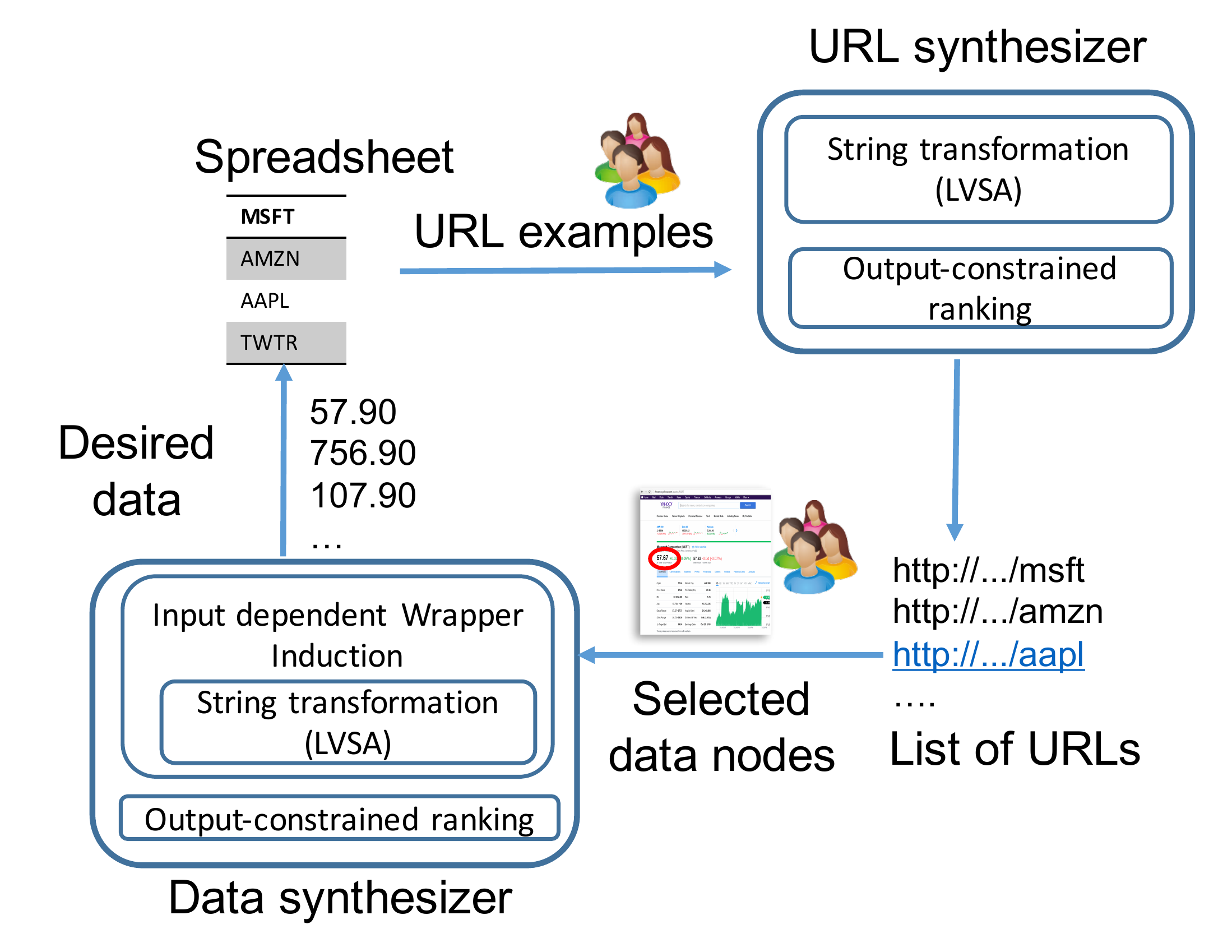}
	\caption{An overview of the workflow of the \sys{} system. A user starts with providing a few examples for URL strings corresponding to the desired webpages of the first few spreadsheet rows. The URL synthesizer then learns a program consistent with the URL examples, which is executed to generate the desired URLs for the remaining spreadsheet rows. The user then opens the URLs for the first few spreadsheet rows in an adjoining pane (one at a time), and highlights the data items that need to be extracted from the webpage. The Data synthesizer then learns a data extraction program consistent with the examples to extract the desired data for the remaining spreadsheet rows.}
	\figlabel{overview}
\end{figure}

$\sys$ is built on top a novel program synthesis paradigm of \quotes{Output-constrained Programming By Examples} (\t{O-PBE}). This formulation is only possible in PBE scenarios where it is possible to compute a finite set of possible outputs for new inputs (i.e. for inputs other than the inputs in the input-output examples provided by the user). Previous PBE systems such as FlashFill~\cite{popl11,cacm12} and its extensions do not have such a property as any output string is equally likely for the new inputs and there is no way to constrain the possible set of outputs. The O-PBE paradigm allows us to develop a new efficient synthesis algorithm that combines both the \emph{output uniqueness constraint} (program should match only 1 output string that is provided by the user for the inputs in the examples) as well as the \emph{generalization constraint} (output is within the set of possible outputs for the other inputs). We instantiate the \t{O-PBE} paradigm for the two sub-tasks in $\sys$: i) URL generation, and ii) input-dependent web extraction. For URL learning problems, the constraint is that the output should be a valid URL or the output should be from the list of possible URLs obtained using the search engine for that particular input. For data extraction problems, the constraint is that the output program for every input should result in a non-empty node in the corresponding HTML document. 

We design an expressive domain-specific language (DSL) for the URL generation programs. The DSL is built on top of regular expression based substring operations introduced in FlashFill~\cite{popl11}. We then present a synthesis algorithm based on \emph{layered version space algebra} (LVSA) to efficiently learn URL programs in the DSL from few input-output examples. We also show that this algorithm is significantly better than existing VSA based techniques. Learning URLs as a string transformation program raises an additional challenge since some URLs may contain additional strings such as unique identifiers that are not in the spreadsheet table and are not constants (\exref{weather} illustrates this challenge). The O-PBE framework allows us to handle such scenarios by having \emph{filter} programs in the language for URLs. These filter programs produce regular expressions for URLs as opposed to concrete strings. Then, $\sys$ leverages a search engine to get a list of relevant URLs for every input and selects the one that matches the regular expression.


Similar to URL learning, $\sys$ uses examples to learn data extraction programs. A user can select a URL to be loaded in a neighboring frame, and then highlight the desired data element to be extracted. $\sys$ records the DOM locations of all such example data and learns a program to perform the desired extraction task for the other rows in the table. We present a new technique for \emph{input-dependent wrapper induction} that involves designing an expressive DSL built on top of XPath constructs and regular expression based substring operations. We, then, present a synthesis algorithm that uses \emph{predicates graphs} to succinctly represent a large number of DSL expressions that are consistent with a given set of I/O examples. For the currency exchange example, our system learns a program that first transforms the date to the required format and then, extracts the conversion rate element whose left sibling contains the transformed date value  from the webpage.

This paper makes the following key contributions:
\begin{itemize}
\item We present $\sys$, a system to join web data with relational data, which divides the data integration task into URL learning and web data extraction tasks.
\item We present a novel program synthesis paradigm called \quotes{Output-constrained Programming By Examples} that allows for incorporating output uniqueness and generalization constraints for an efficient synthesis algorithm (\secref{problem}). We instantiate this paradigm for the two sub-tasks of URL learning and data extraction. 
\item We design a DSL for URL generation using string transformations and an algorithm based on \emph{layered version space algebra} to learn URLs from a few examples (\secref{url-learn}).
\item We design a DSL on top of XPath constructs that allows input-dependent data extractions. We present a synthesis algorithm based on a \emph{predicates graph} data structure to learn extraction programs from examples (\secref{data-ext}).
\item We evaluate $\sys$ on 88 real-world data integration tasks. It takes on average less than 1.2 examples and 0.15 seconds each to learn the URLs, whereas it takes less than 5 seconds and 1 example to learn data extraction programs for 93\% of tasks (\secref{eval}).
\end{itemize}


\section{Motivating Examples}\seclabel{motivation}

In this section, we present a few additional motivating scenarios for web data integration tasks of varying complexity and illustrate how \sys{} can be used to automate these tasks using few input-output examples. 

\begin{example}\exlabel{stock}
\textbf{[Stock prices]} A user wanted to retrieve the stock prices for hundreds of companies (\t{Company} column) as shown in \figref{ex-stock}. 

\end{example}

\begin{figure}[!htpb]
\begin{tabular}{c c}
\begin{minipage}{0.6\linewidth}
\small
	\begin{center}
		\begin{tabular}{|c|l|l|c|}
			\hline
			& Company & \multicolumn{1}{c}{URL} & Stock price \\\hline\hline
			1 &  MSFT  & https://finance.yahoo.com/q?s=msft & 59.87\\\hline
			2 & AMZN & \textbf{https://finance.yahoo.com/q?s=amzn} & \textbf{775.88} \\\hline
			3 & AAPL  & \textbf{https://finance.yahoo.com/q?s=aapl} & \textbf{113.69}\\\hline
			4 & TWTR  & \textbf{https://finance.yahoo.com/q?s=twtr} & \textbf{17.66}\\\hline
			5 & T  &  \textbf{https://finance.yahoo.com/q?s=t} & \textbf{36.51}\\\hline
			6 & S  & \textbf{https://finance.yahoo.com/q?s=s} & \textbf{6.31}\\\hline
		\end{tabular}
	\end{center}

\end{minipage}
&
\begin{minipage}{0.4\linewidth}
\centering
\includegraphics[scale=0.31]{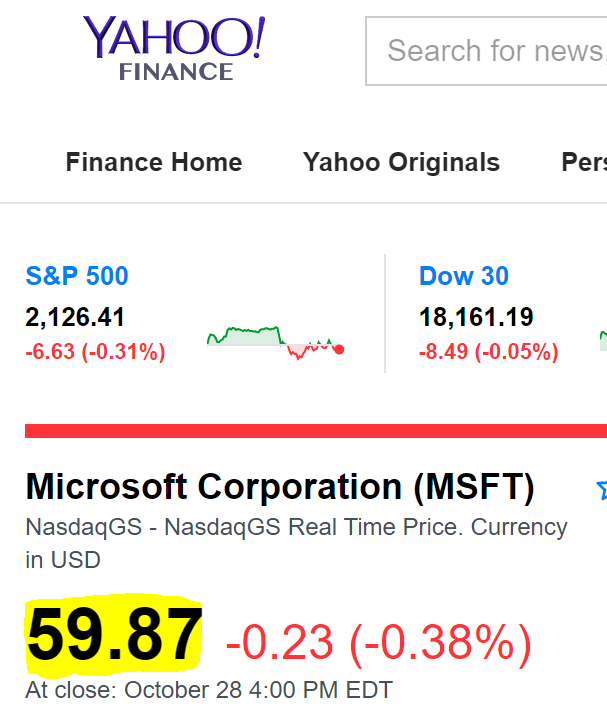}
\end{minipage}
\\
(a) & (b)
\end{tabular}
	\caption{Joining stock prices from web with company symbols using \sys{}. Given one example URL and data extraction from the webpage for the first row, the system learns a program to automatically generate the corresponding URLs and extracted stock prices for the other rows (shown in bold).}
	\figlabel{ex-stock}
\end{figure}

In order to perform this integration task in \sys{}, a user can provide an example URL such as \url{https://finance.yahoo.com/q?s=msft} that has the desired stock price for the first row in the spreadsheet (\figref{ex-stock}(a)). This web-page then gets loaded and is displayed to the user. The user can then highlight the required data from the web-page (\figref{ex-stock}(b)). \sys{} learns the desired program to perform this task in two steps. First, it learns a program to generate the URLs for remaining rows by learning a string transformation program that combines the constant string \quotes{https://finance.yahoo.com/q?s=} with the company symbol in the input. Second, it learns a web data extraction program to extract the desired data from these web-pages.




\begin{example}\exlabel{weather}
\textbf{[Weather]}. A user had a list of addresses in a spreadsheet and wanted to get the weather information at each location as shown in \figref{ex-weather}. The provided example URL is \url{https://weather.com/weather/today/l/Seattle+WA+98109:4:US#!}.  
\end{example}

\begin{figure}[!htpb]
\begin{tabular}{c c}
\begin{minipage}{0.5\linewidth}
	\small
	\begin{center}
		\begin{tabular}{|c|c|c|}
			\hline
			& Address & Weather \\\hline\hline
			1 &742 17th Street NE,Seattle,WA
			 & 59\\\hline
			2 &732 Memorial Drive,Cambridge,MA
			 & \textbf{43} \\\hline
			3 & Apt 12, 112 NE Main St.,Boston,MA
			 & \textbf{42} \\\hline

		\end{tabular}
	\end{center}

\end{minipage}
&
\begin{minipage}{0.5\linewidth}
\centering
\includegraphics[scale=0.25]{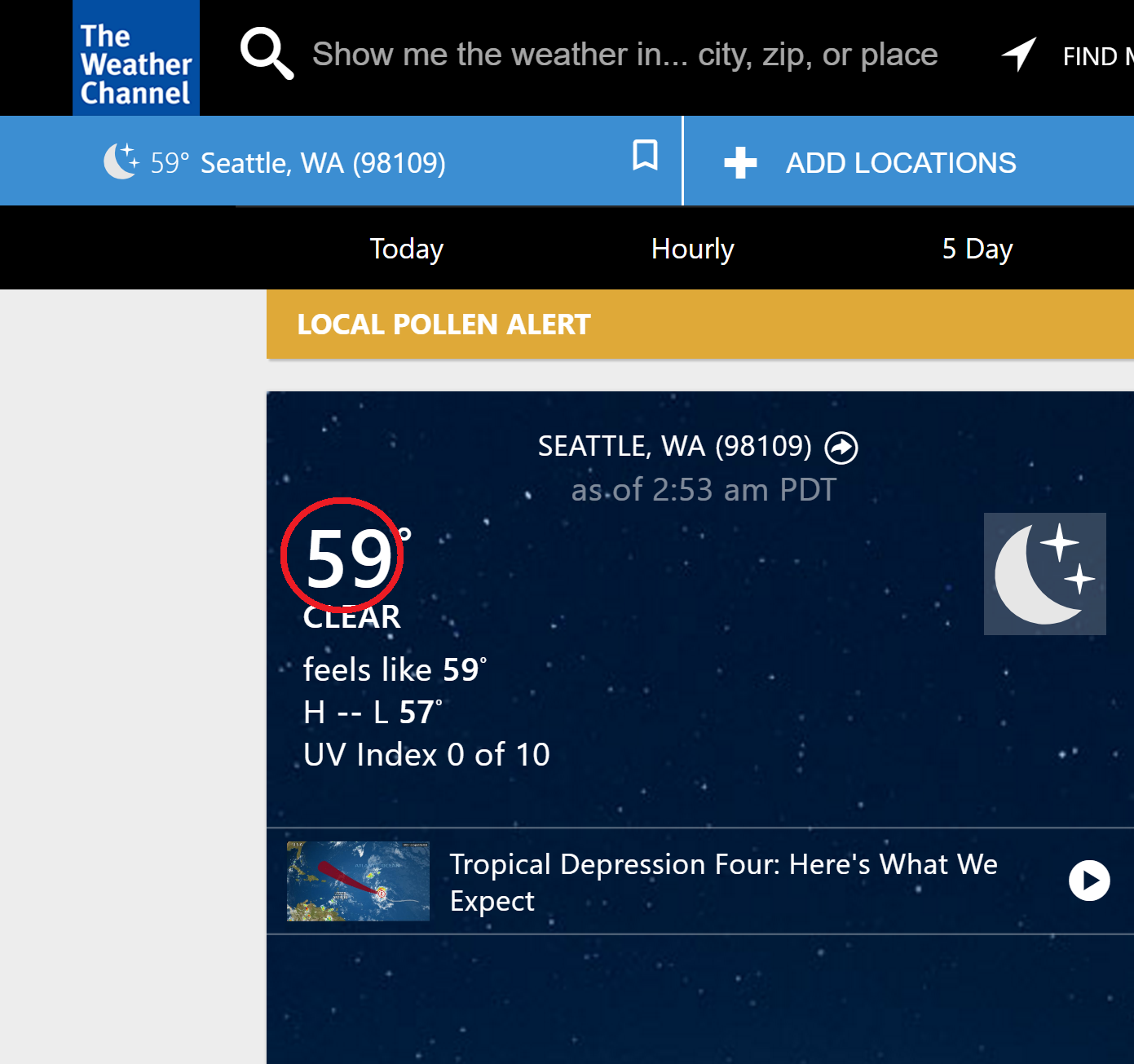}
\end{minipage}
\\
(a) & (b)
\end{tabular}
	\caption{Joining weather data from web with addresses. Given one example row, the system automatically extracts the weather for other row entries (shown in bold).}
	\figlabel{ex-weather}
\end{figure} 

There are two challenges in learning the URL program for this example. First, the addresses contain more information than just the city and state names such as street names and house numbers. 
Therefore, the URL  program first needs to learn regular expressions to extract the city-name and state from the address and then concatenate them appropriately with  constant strings to get the desired URL. The second challenge is that the URL contains zip code that is not present in the input, meaning that there is no simple program that concatenates constants and sub-strings to learn the URLs for the remaining inputs. For supporting such cases, the DSL for URL learning also supports \emph{filter} programs that use regular expressions to denote unknown parts of the URL. A possible \emph{filter} program for this example is
https://weather.com/weather/today/l/\{\textbf{Extracted city name}\}+\{\textbf{Extracted state name}\}+\{\textbf{AnyStr}\}:4:US\#!, where \textbf{AnyStr} can match any non-empty string. Then, for every other row in the table, $\sys$ leverages a search engine to get a list of possible URLs for that row and selects the top URL that matches the filter program. By default, we use the words in input row as the search query term and set the target URL domain to be the domain of the given example URLs. \sys{} also allows users to provide search query terms (such as ``Seattle weather'') as additional search query examples and it learns the remaining search queries for other inputs using a string transformation program. Using the search query and a filter program, \sys{} learns the following URL for the second row: \url{https://weather.com/weather/today/l/Cambridge+MA+02139:4:US#!}.

\begin{example}\textbf{[Citations]} A user had a table containing author names and titles of research papers, and  wanted to extract the number of citations for each article using Google Scholar (as shown in \figref{ex-gsch}).
\end{example}  

In this case, the example URL for Samuel Madden's Google Scholar page is \url{https://scholar.google.com/scholar?q=samuel+madden} and the corresponding web page is shown in \figref{ex-gsch}(b). The URL can be learned as a string transformation program over the \t{Author} column. The more challenging part of this integration task is that the data in the web page is in a semi-structured format and the required data (\# citations) should be extracted based on the \t{Article} column in the input. 
 Our data extraction DSL is expressive enough to learn a program that captures this complex dependency. This program generates the entire string ``Cited by 2316'' and $\sys$ learns another transformation program  on top of this to extract only the number of citations ``2316'' from the results.

\begin{figure}[!htpb] 
\begin{tabular}{c c}
\begin{minipage}{0.5\linewidth}
		\begin{adjustbox}{max width=\linewidth}
			\begin{tabular}{|c|c|c|c|}
				\hline
				& Author  & Article & $\#$ citations \\\hline\hline
				1 & Samuel Madden & TinyDB: an acquisitional ... &  2316 \\ \hline
				2 & HV Jagadish & Structural joins: A primitive ... & \textbf{1157} \\ \hline 
				3 & Mike Stonebraker & C-store: a column-oriented ... & \textbf{1119} \\ \hline
			\end{tabular}
\end{adjustbox}
		
\end{minipage}
&
\begin{minipage}{0.5\linewidth}
\includegraphics[scale=0.25]{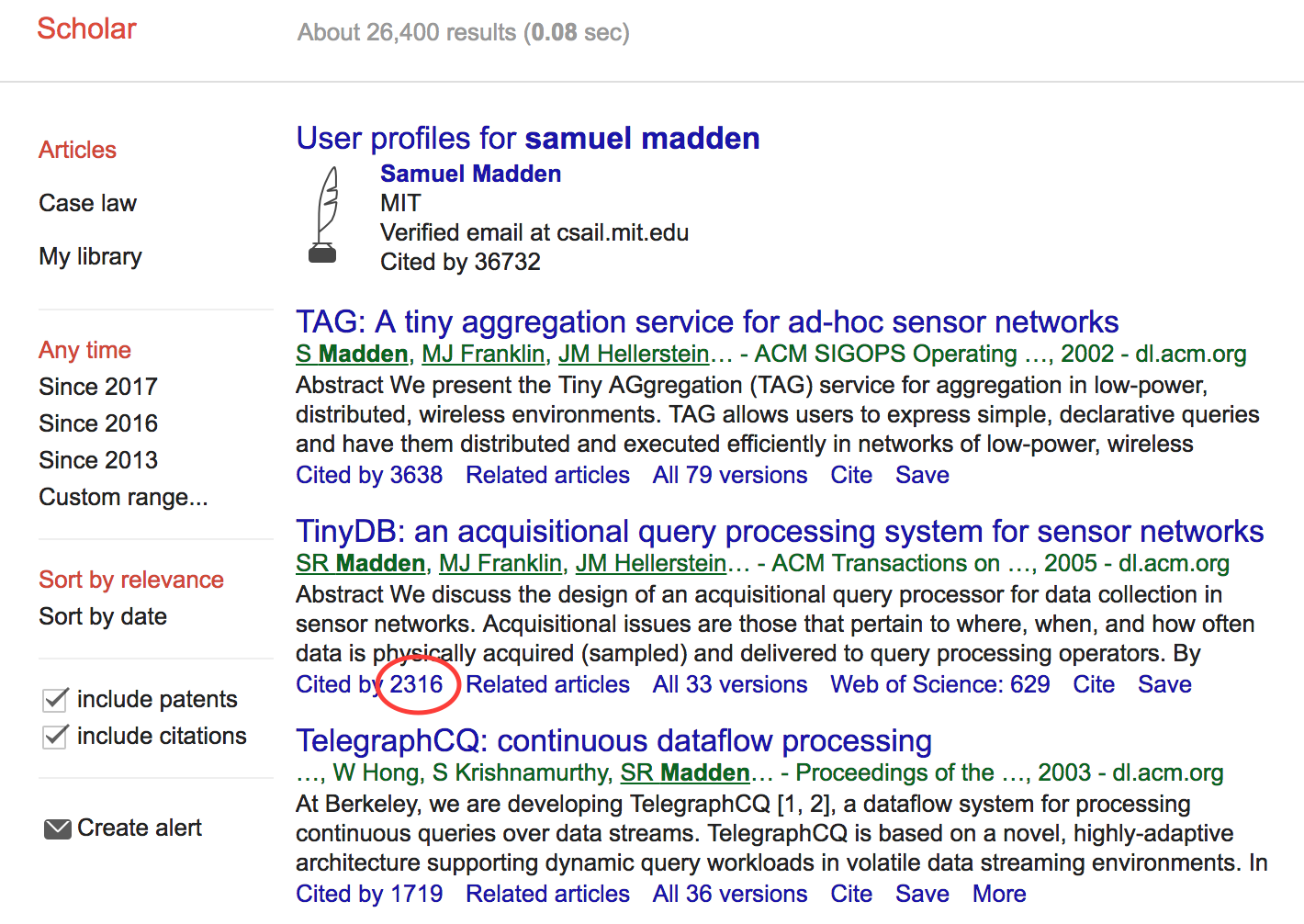}
\end{minipage}

\\
(a) & (b)

\end{tabular}
	\caption{(a) Integrating a spreadsheet containing authors and article titles with the number of citations obtained from Google Scholar. (b) The google scholar page for the first example.}
	\figlabel{ex-gsch}
	
\end{figure} 

\section{Output-constrained Programming By Example}\seclabel{problem}
We first define the abstract \emph{Output-constrained} Programming By Example (O-PBE) problem, which we then instantiate for both the URL synthesizer and the data extraction synthesizer. Let $E = \{(i_1, o_1), \cdots, (i_m, o_m) \}$ denote the list of $m$ input-output examples provided by the user and $U = \{i_{m+1}, \cdots, i_{m+n}\}$ denote the list of  $n$ inputs with unknown outputs. We use $L$ to denote the DSL that describes the space of possible programs. The traditional PBE problem is to find a program $P$ in the DSL that satisfies the given input-output examples i.e. $$\exists~ P \in L.~\forall~k \leq m.~P(i_k) = o_k$$
On the other hand, the O-PBE problem formulation takes advantage of the existence of an \emph{oracle} $O$ that can generate a finite list of possible outputs for any given input. For instance, in the URL learning scenario, this oracle is a search engine that lists the top URLs obtained for a search query term based on the input. For the data extraction learning scenario, this oracle is the web document corresponding to each input where any node in the web document is a candidate output. The existence of this oracle can benefit the PBE problem in two ways. First, we can solve  problems with noisy input-output examples where there is no single  program in the language that can \emph{generate} all the desired outputs for the inputs in the examples. For instance, the URL learning  task in \exref{weather} is a synthesis problem with noisy input-output examples that traditional PBE approaches cannot solve. However, the presence of oracles can solve this problem because it is now sufficient to just learn a program that, given an input and list of possible outputs, can discriminate the desired output from the other possible outputs. This property is called the \emph{output-uniqueness constraint}.  
The second benefit of oracles is that they impose additional constraints on the learned program. In addition to satisfying the input-output examples, we want the learned program to produce valid outputs for the other inputs. We refer to this as the \emph{generalization constraint}. Using this constraint, the O-PBE approach can efficiently learn the desired program using very few input-output examples. Thus, we can now define the O-PBE problem formally as follows:
\begin{eqnarray*}
\exists~P \in L.~\forall ~k \leq m.~P(i_k, O(i_k)) = o_k & \textit{(output-uniqueness constraint)}\\
~\wedge ~ \forall~k \leq n. ~P(i_{m+k}, O(i_{m+k})) \in O(i_{m+k}) & \textit{(generalization constraint)}
\end{eqnarray*}

where program $P$ is now a more expressive higher-order expression in the language $L$ that takes as input a list of possible outputs $O(i)$ (in addition to the input $i$) and returns one output among them i.e. $P(i, O(i))  = o~ s.t.~o \in O(i)$. 
%



At a high level, to solve this synthesis problem, \sys{} first learns the set of all programs in the language $L$ that satisfies the given input-output examples $E$ (not necessarily uniquely). This set is represented succinctly using special data structures. Then, \sys{} uses an \emph{output-constrained ranking scheme} to eliminate programs that do not uniquely satisfy the given examples $E$ or are inconsistent with the unseen inputs $U$. 

We now describe the two instantiations of the abstract O-PBE problem for the URL and data extraction synthesizers in more detail.
\section{URL Learning}
\seclabel{url-learn}

We first present the domain-specific language for the URL generation programs and then present a synthesis algorithm based on \emph{layered version space algebra} to efficiently learn programs from few input-output examples.

\begin{figure*}[t]
\small
\centering
\begin{eqnarray*}
\mbox{URL String } u & := &  \h{\t{Filter}($\phi$)} \nonumber \\
\mbox{Predicate } \phi & := & \t{Concat}(f_1, \cdots, f_n) \nonumber \\
\mbox{Atomic expr } f & := & r \; \vert \; b \nonumber \\
\h{\mbox{Regex expr } r} & := & \h{\t{AnyStr}}  \nonumber \\
\mbox{Base expr } b & := & \t{ConstStr}(s) \; \vert \; \t{SubStr}(k,p_l,p_r,c) \; \vert \; \h{\t{Replace}($k, p_l,p_r,c, s_1,s_2$)} \nonumber \\
\mbox{Position } p & := & (\tau, k, \t{Dir}) \; \vert \; \t{ConstPos}(k)  \\
\mbox{Direction } \t{Dir} & := & \t{Start} \; \vert \; \t{End} \nonumber \\
\mbox{Token } \tau & := & s \; \vert \; T\nonumber \\
\mbox{Regex Token } T & := & \t{CAPS} \; \vert \; \t{ProperCase} \; \vert \; \t{lowercase} \; \vert \; \t{Digits} \; \vert \; \t{Alphabets} \; \vert \; \t{AlphaNum}\nonumber \\
\mbox{Case } c & := & \t{lower} \; \vert \; \t{upper} \; \vert \; \t{prop} \; \vert \; \t{iden}
\end{eqnarray*}
\caption{ The syntax of the DSL $\lurl$ for regular expression based URL learning. Here, $s$ is a string,  and $k$ is an integer.}
\figlabel{urldslsyntax}
\end{figure*}

\subsection{URL Generation Language $\lurl$}

\paragraph*{Syntax} The syntax of the DSL for URL generation $\lurl$ is shown in \figref{urldslsyntax}. The DSL is built on top of regular expression based substring constructs introduced in FlashFill~\cite{popl11,cacm12} with the key differences  highlighted. The top-level URL program is a filter expression $\t{Filter}(\phi)$ that takes as argument a predicate $\phi$, where $\phi$ is denoted using a concatenation of base atomic expressions or \t{AnyStr} expressions.  The base atomic  expression $b$ can either be constant string, a regular expression based substring expression that takes an index, two position expressions and a case expression, or a replace expression that takes two string arguments in addition to the arguments of substring expression. A position expression can either be a constant position index $k$ or a regular expression based position $(\tau,k,\t{Dir})$ that denotes the \t{Start} or \t{End} of $k^\t{th}$ match of token $\tau$ in the input string.

\paragraph*{Semantics} The semantics of the DSL for $\lurl$ is shown in \figref{urldslsemantics}. We use the notation $\sem{x}$ to represent the semantics of $x$ when evaluated on an input $i$ (a list of strings from the table row).
The semantics of a filter expression $\t{Filter}(\phi)$  is to use a URL list generator oracle $O_u$ to obtain a ranked list of URLs for the input $i$, and select the first URL that matches the regular expression generated by the evaluation of the predicate $\phi$. The default implementation for $O_u$ runs a search engine on the words derived from the input $i$ (with the domain name of URL examples) and returns the URLs of the top results. However, a user can also provide a few search query examples, which \sys{} uses to learn another string transformation  program to query the search engine for the remaining rows. The semantics of 
a predicate expression $\phi$ is to first evaluate each individual atomic expression in the arguments and then return the concatenation of resulting atomic strings. The semantics of \t{AnyStr} expression is $\Sigma^+$ that can match any non-empty string. The semantics of a substring expression is to first evaluate the position expressions to obtain left and right indices and then return the corresponding substring for the $k$th string in $i$. The semantics of a replace expression $\t{Replace}(k,p_l,p_r,c,s_1,s_2)$ is to first get the substring corresponding to the position expressions and then replace all occurrences of string $s_1$ with the string $s_2$ in the substring. We allow strings $s_1$ and $s_2$ to take values from a finite set of delimiter strings such as `` '', ``-'', ``\_'',  and ``\#''. 

\begin{figure*}[t]
\small
\centering
\begin{eqnarray*}
\sem{\t{Filter}(\phi)} & = & u~ \mbox{s.t. } u \in O_u(i) \mbox{ and } \sem{\phi} \models u \\
\sem{\t{Concat}(f_1,\cdots,f_n)} & = & \t{Concat}(\sem{f_1}, \cdots, \sem{f_n}) \\
\sem{\t{AnyStr}} & = & \Sigma^+ \\
\sem{\t{ConstStr}(s)} & = & s \\
\sem{\t{SubStr}(k, p_l,p_r,c)} & = & \t{ToCase}(i_k[\sems{p_l}..\sems{p_r}],c)\\
\sem{\t{Replace}(k,p_l,p_r,c,s_1,s_2)} & = & \t{ToCase}(i_k[\sems{p_l}..\sems{p_r}],c)[s_1 \leftarrow s_2]\\
\sem{\t{ConstPos}(k)} & = & k > 0 ? \; k : \t{len}(s)+k \\
\sem{(\tau,k,\t{Start})} & = & \mbox{Start of } k^{\t{th}} \mbox{match of $\tau$ in i}\\
\sems{(\tau,k,\t{End})} & = & \mbox{End of } k^{\t{th}} \mbox{match of $\tau$ in i}
\end{eqnarray*}

\caption{The semantics of the DSL $\lurl$. $O_u$ is a URL list  oracle that generates a ranked list of URLs for the input $i$ by using a search engine.}
\figlabel{urldslsemantics}
\end{figure*}

\paragraph*{Examples} A program in $\lurl$ to perform the URL learning task in \exref{cur} is:
\t{Filter}(\t{Concat}(\\\t{ConstStr}(\quotes{http://www.investing.com/currencies/}), \t{Substr}(0, \t{ConstPos}(0), \t{ConstPos}(-1), \t{lower}), \t{ConstStr}(\quotes{-}), \t{Substr}(1, \t{ConstPos}(0), \t{ConstPos}(-1), \t{lower}), \t{ConstStr}(\quotes{-historical-data}))).
 The program concatenates a constant string, the lowercase transformation of the first input string ($-1$ index in \t{ConstPos} denotes the last string index), the constant hyphen , the lowercase transformation of the second input string, and finally, another constant string.
 
 A DSL program for the URL learning task in \exref{weather} is: \t{Filter}($\phi$), where $\phi\equiv~$\t{Concat}($b_1$, $b_2$, \t{ConstStr}(\quotes{+}), $b_3$,\t{ConstStr}(\quotes{+}), \t{AnyStr}, \t{ConstStr}(\quotes{:4:US\#!})),
 $b_1\equiv~$\t{ConstStr}(\quotes{https://weather.com\\/weather/today/l/}), $b_2\equiv$\t{SubStr}(0, (\quotes{,},-2,\t{End}),(\quotes{,},-1,\t{Start}),\t{iden}),  and $b_3\equiv~$\t{SubStr}(0, (\quotes{,},-1,\t{End}),\\\t{ConstPos}(-1),\t{iden}). Here, $b_2$ and $b_3$ are regular expression based substring expressions to derive the city and the state strings from the input address.

\subsection{Synthesis Algorithm}
We now present the synthesis algorithm to learn a URL program that solves the O-PBE problem. 

\subsubsection{Background: Version Space Algebra}
This section presents a brief background on version space algebra (VSA), a technique used by existing string transformation synthesizers such as FlashFill. We refer the readers to ~\cite{cacm12} for a detailed description. The key idea of VSA is to use a directed acyclic graph (DAG) to succinctly represent an exponential number of candidate programs using polynomial space. For the string transformation scenario, a DAG $\mathcal{D}$ is defined as a tuple $(\mathcal{V}, \nu_s, \nu_t, \mathcal{E}, \mathcal{W})$ where $\mathcal{V}$ is the set of vertices, $\mathcal{E}$ is the set of edges, $\nu_s$  is the starting vertex and $\nu_t$ is the target vertex. Each edge in the DAG represents a set of atomic expressions (the map $\mathcal{W}$ captures this relation) whereas a path in the DAG represents the concatenation of the atomic expressions of the edges. Given a synthesis problem, a DAG is constructed in such a way that any path in the DAG from $\nu_s$ to $\nu_t$ is a valid program in $L$ that satisfies the examples. This is achieved by iteratively constructing a DAG for each example and performing an automata-like-intersection on these individual DAGs to get the final DAG. For example, a sample DAG is shown in \figref{dag}. The nodes in this DAG (for each I/O example) correspond to the indices of the output string. An edge from a node $i$ to a node $j$ in the DAG represents the set of expressions that can generate the substring between the indices $i$ and $j$ of the output example string when executed on the input data.

\begin{figure}
	\centering
	\includegraphics[scale=0.18]{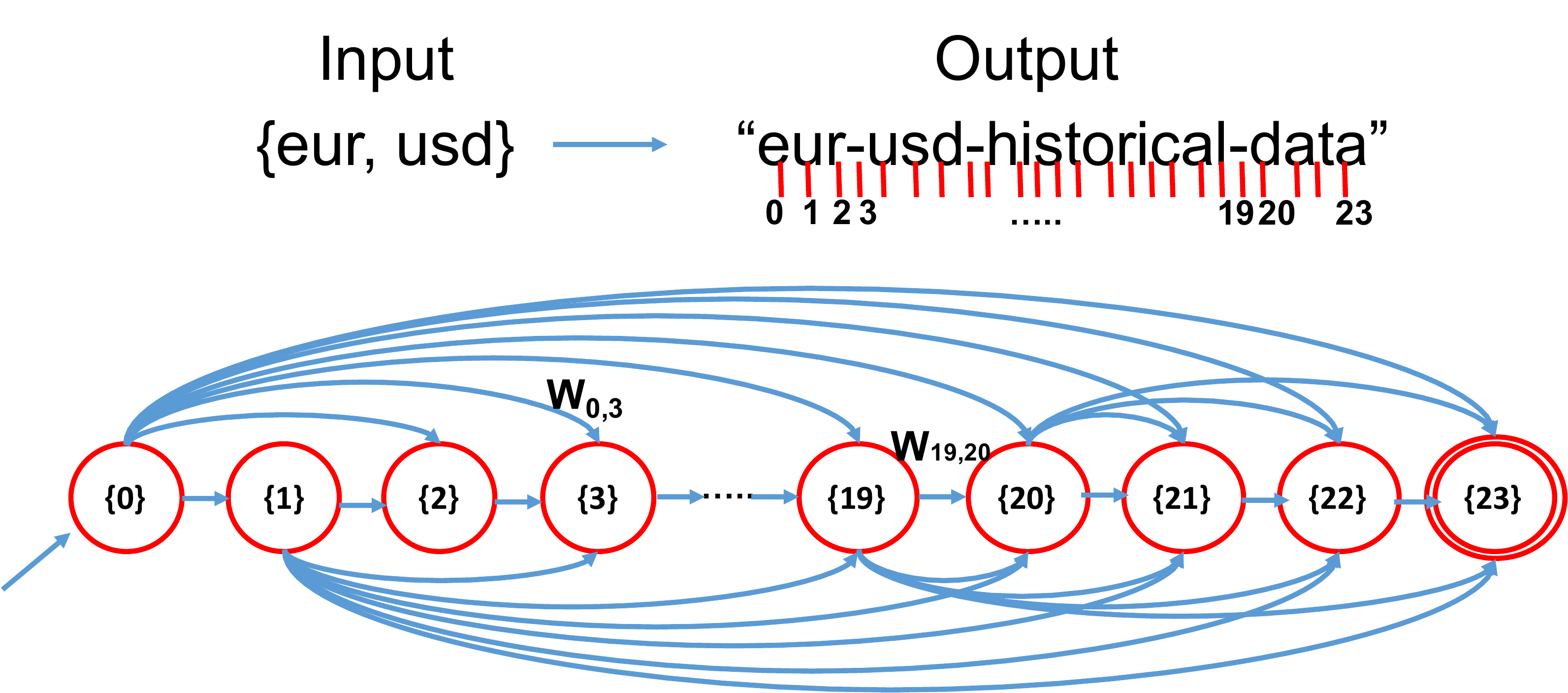}
	\caption{An example DAG using Version space algebra (VSA) to represent the set of transformations consistent with the example. For example, here, $~\mathcal{W}_{0,3}$  is a list of three different expressions  $\{\t{ConstStr}(\t{eur}),~ \t{SubStr}(0, \t{ConstPos}(0), \t{ConstPos}(-1), \t{lower}),$ $\t{AnyStr}\}$; $\mathcal{W}_{19,20} =\{\t{ConstStr}(\t{d}), ~\t{SubStr}(1, \t{ConstPos}(2),$ $ \t{ConstPos}(3),\t{lower}), ~\t{AnyStr} \}$.}
	\figlabel{dag}
\end{figure}

\subsubsection{Layered Version Space Algebra}\seclabel{url}

It is challenging to use VSA techniques for learning URLs because the URL strings are long, and the run time and the DAG size of the existing algorithms explode with the length of the output.
To overcome this issue, we introduce a synthesis algorithm based on a layered version space algebra for efficiently searching the space of consistent programs. The key idea is to perform search over increasingly expressive sub-languages $L_1 \subseteq L_2 \cdots \subseteq L_k$, where $L_k$ corresponds to the complete language $\lurl$. 
The sub-languages  are selected such that the earlier languages capture the portion of the search space that is highly probable and at the same time, it is easier to search for a program in these sub-languages. For example, it is less likely to concatenate a constant with a sub-string within a word in the URL and hence, earlier sub-languages are designed to eliminate such programs. For instance, consider the URL in \exref{cur}. In this case, given the first I/O example, programs that derive the character \t{d} in \t{data} from the last character in the input (\t{USD}) are not desirable because they do not generalize to other inputs. 

The general synthesis algorithm $\GenProg$ for learning URL string transformations in a sub-language $L_i$ is shown in \figref{gensearch}. This algorithm is similar to the VSA based algorithm, but the key difference is that instead of learning all candidate programs (which can be expensive), the algorithm only learns the programs that are in the sub-language $L_i$. The $\GenProg$ algorithm takes as input the I/O examples $\{i_k, o_k\}_k$, and three Boolean functions $\substrlambda$, $\constlambda$, $\randomlambda$ that parameterize the search space for the language $L_i$. The output is a program that is consistent with the examples or $\bot$ if no such program exists. The algorithm first uses the $\GenDag$ procedure to learn a DAG consisting of all consistent programs (in the given sub-language) for the first example. It then iterates over the other examples and intersects the corresponding DAGs to compute a DAG representing programs consistent with all the examples. Finally, it searches for the best program in the DAG that satisfies the O-PBE constraints and returns it. 

The $\GenDag$ algorithm is also shown in \figref{gensearch}, where the space of programs is constrained by the parameter Boolean functions $\substrlambda$, $\constlambda$, and $\randomlambda$. Each function $\lambda:\t{int}\rightarrow\t{int}\rightarrow\t{string}\rightarrow\t{bool}$ takes two integer indices and a string as input and returns a Boolean value denoting whether certain atomic expressions are allowed to be added to the DAG.  The algorithm first creates $\t{len}(\entityo)+1$ nodes, and then adds an edge between each pair of nodes $\langle k, l \rangle$ such that $0\leq k < l \leq \t{len}(\entityo)$.  For each edge $\langle k,l \rangle$, the algorithm learns all atomic expressions that can generate the substring $\entityo[k..l]$. For learning \t{SubStr} and \t{Replace} expressions, the algorithm enumerates different argument parameters for positions, cases, and delimiter strings, whereas the \t{ConstStr} and \t{AnyStr} expressions are always available to be added. The addition of \t{SubStr} and \t{Replace} atomic expressions to the DAG are guarded by the Boolean function $\substrlambda$ whereas the addition of \t{ConstStr} and \t{AnyStr} atomic expressions are guarded by the Boolean functions  $\constlambda$ and $\randomlambda$, respectively.

\begin{figure}[t]
	\small
	\begin{tabular}{c c}
		\begin{minipage}{0.4\linewidth}
			\begin{program}[style=tt, number=false]
				\stab{1mm}\underline{$\GenProg$}($\{(\entityi_k, \entityo_k)\}_k$, $\substrlambda$, $\constlambda$, $\randomlambda$)
				$\Dag$ d = $\GenDag$($\entityi_1$,$\entityo_1$,$\substrlambda$,$\constlambda$,$\randomlambda$)
				for\tab ~k from 2 to $m$:
				if  d = $\bot$: return $\bot$
				$\Dag$ d' = $\GenDag$($\entityi_k$,$\entityo_k$,$\substrlambda$,$\constlambda$,$\randomlambda$)
				d = d.$\Intersect$(d')
				\untab
				return \t{SearchBestProg}(d)
			\end{program}
		\end{minipage}
		&
		\begin{minipage}{0.6\linewidth}
			\begin{program}[style=tt, number=false]
				\stab{1mm}\underline{\GenDag}($\entityi$, $\entityo$, $\substrlambda$, $\constlambda$, $\randomlambda$)
				$\nodes$ = \{0, ..., $\len$(o)\}, $\nodestart$ = \{0\}, $\nodetarget$ = \{$\len$(o)\}
				$\edges$  = \{$\langle k, l \rangle$ : 0 $\leq$ k < l $\leq$ $\len$(o)\}
				$\mathcal{W}$ = maps each edge to set of atomic exprs
				for\tab each $0 \leq k < l \leq \len(\entityo)$: 
				$\edgemap$ = $\emptyset$
				if~\tab $\substrlambda(k,l,o)$: $\edgemap$.$\add$($\GenSubstr$($k,l,$ $\entityi$, $\entityo$)) 
				\untab
				if~\tab $\substrlambda(k,l,o)$: $\edgemap$.$\add$($\t{GenReplace}$($k,l,$ $\entityi$, $\entityo$)) 
				\untab
				if~\tab $\constlambda(k,l,o)$: $\edgemap$.$\add$($\t{ConstStr}(\entityo[k..l])$)
				\untab
				if~\tab $\randomlambda(k,l,o)$: $\edgemap$.$\add$($\t{AnyStr}$)
				\untab
				$\mathcal{W}$.add($\langle k, l\rangle$,~$\edgemap$)
				\untab
				return $\Dag$($\nodes$, $\nodestart$, $\nodetarget$, $\edges$, $\mathcal{W}$)
			\end{program}
			
		\end{minipage}
	\end{tabular}
	\caption{Synthesis algorithm for URL generation programs, parameterized by $\substrlambda$, $\constlambda$, and $\randomlambda$.}
	\figlabel{gensearch}
\end{figure}


\begin{figure}[t]
	\small
	\begin{program}[style=tt, number=false]
		\stab{1mm}\underline{$\t{LearnURL}$}($\{(\entityi_k, \entityo_k)\}_k$)
		if($p := \GenProg(\{(\entityi_k, \entityo_k)\}_k,\t{oW},\t{oW},\t{oW}) \neq \bot$: return $p$ // Layer 1
		if($p := \GenProg(\{(\entityi_k, \entityo_k)\}_k,\t{mW},\t{oW},\t{oW}) \neq \bot$: return $p$ // Layer 2
		if($p := \GenProg(\{(\entityi_k, \entityo_k)\}_k,\t{iW} \lor \t{mW},\t{oW},\t{oW}) \neq \bot$: return $p$ // Layer 3
		return $\GenProg(\{(\entityi_k, \entityo_k)\}_k,\t{T},\t{T},\t{T})$ // Layer 4
	\end{program}
	\caption{Layered version spaces for learning a program in  $\lurl$.}
	\figlabel{learnealgo}
\end{figure}

\figref{learnealgo} shows how the different layers are instantiated for learning URL expressions. 
For the first layer, the algorithm only searches for URL expressions where each word in the output is either generated by a substring or a constant or an \t{AnyStr}. The \t{onlyWords} (\t{oW}) function is defined as:
\begin{equation*}
\t{oW} = (i,j,\entityo) => \neg\t{isAlpha}(\entityo[i-1]) \land  \neg\t{isAlpha}(\entityo[j+1]) \land \forall~k: i\leq k \leq j ~ \t{isAlpha}(\entityo[k])
\end{equation*}
where \t{isAlpha(c)} is true if the character \t{c} is an alphabet. 
The second layer allows for multiple words in the output string to be learned as a single substring. The function \t{multipleWords} (\t{mW}) is defined as:
\begin{equation*}
\t{mW} = (i,j,\entityo) => \neg\t{isAlpha}(\entityo[i-1]) ~ \land ~ \neg\t{isAlpha}(\entityo[j+1])
\end{equation*}

The third layer, in addition, allows for words in the output string to be a concatenation of multiple substrings, but not a concatenation of substrings with constant strings (or \t{AnyStr}). The function \t{insideWords} (\t{iW}) is defined as:

\begin{equation*}
\t{iW} = (i,j,\entityo) => \forall~k: i\leq k \leq j ~ \t{isAlpha}(\entityo[k])
\end{equation*}

The final layer allows  arbitrary compositions of constants, \t{AnyStr} and substring expressions by setting the functions $\substrlambda$, $\constlambda$ and $\randomlambda$ to always return \t{True} (\t{T}).

\begin{example}
 Consider the currency exchange example where the example URL for the input \{\t{EUR}, \t{USD}, \t{03, November 16}\} is \quotes{\url{http://www.investing.com/currencies/eur-usd-historical-data}}. The first layer of the search will create a DAG as shown in \figref{hsearchdag}. We can observe that the DAG eliminates most of the unnecessary programs such as those that consider the \t{d} in \t{data} to come from the \t{D} in \t{USD} and the DAG is much smaller (and sparser) compared to the DAG in \figref{dag}.
\end{example}

\begin{figure}
\centering
\includegraphics[scale=0.20]{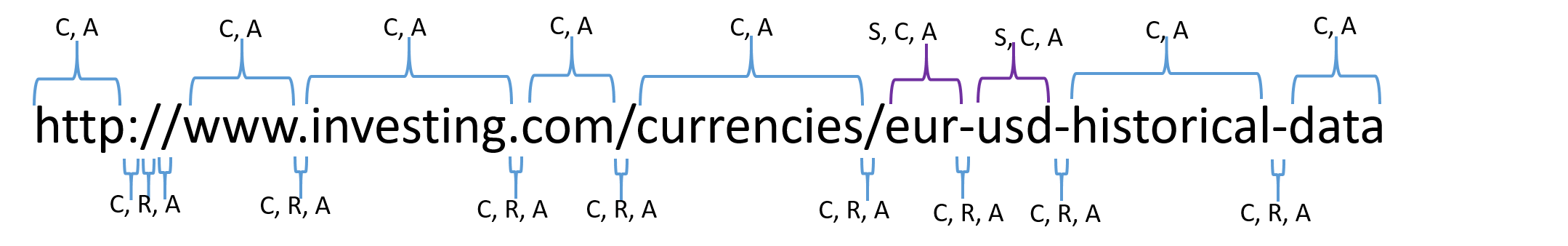}
\caption{DAG for \exref{cur} constructed using layer 1 where S: \t{SubStr}, C: \t{ConstStr}, R: \t{Replace}, and A: \t{AnyStr}. }
\figlabel{hsearchdag}
\end{figure}

\begin{example}
For the same example, assume that we want to get the currency exchange values from \url{http://finance.yahoo.com/q?s=EURUSD=X}. For this example, layers 1 and 2 can only learn the string \t{EURUSD} as a \t{ConstStr} which will not work for the other inputs, or a \t{AnyStr} which is too general. So, the layered search moves to layer 3 which allows \t{SubStr} inside words. Now, the system can learn \t{EUR} and \t{USD} separately as two \t{SubStr} expressions and concatenate them.
\end{example}

\begin{example}\exlabel{country}
Consider another example where we want to learn the URL \url{https://en.wikipedia.org/wiki/United\_States} from the input \t{United States} and the URL \url{https://en.wikipedia.org/wiki/India} from the input \t{India}. \figref{hsearchdag2}(a) and (b) show portions of the DAGs for these two examples when using the first layer\footnote{We omit the DAGs for the first part of the URLs as they are similar to the previous example.}.  Here, there is no common program that can learn both these examples together. In such situations, the layered search will move to the second layer. \figref{hsearchdag2}(c) shows the extra edges added in layer 2. Now, these examples can be learned together as \t{Filter}(\t{Concat}(\t{ConstStr}(\quotes{https}),$\cdots$,\t{ConstStr}(\quotes{/}), \t{Replace}(0, \t{ConstPos}(0),\t{ConstPos}(-1), \t{iden}, \quotes{ }, \quotes{\_}))).
\end{example}

\begin{figure}
	\includegraphics[scale=0.25]{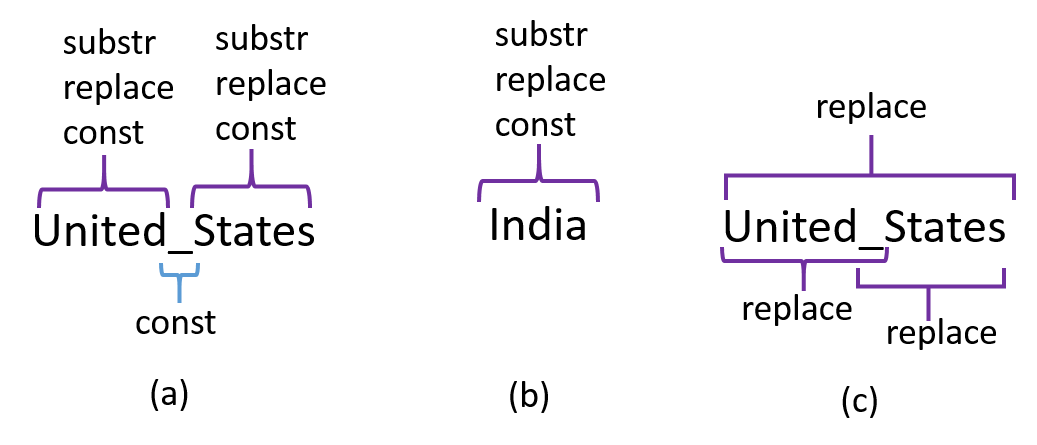}
	\caption{(a) and (b) shows a portion of the DAGs for \exref{country} using layer 1 of hierarchical search. (c) shows additional edges from layer 2 of the hierarchical search.}
	\figlabel{hsearchdag2}
\end{figure}



\subsubsection{Output-constrained Ranking}
The \t{LearnURL} algorithm learns multiple programs that match the example URLs.
 However, not all of these programs are desirable for two reasons. First, some filter programs are too general and hence, fail the output-uniqueness constraint.  For instance, \t{AnyStr} is one of the possible predicates that will be learned by the algorithm, but in addition to matching the desired example URLs, this predicate will also match any URL in the search results. Hence, we need to carefully select a consistent program from the DAG.  Second, not all programs are equally desirable as they may not generalize well to the unseen inputs. 
  For instance, consider \exref{weather}. If we have an example URL such as \url{https://weather.com/weather/today/l/Seattle+WA+98109:4:US#!}, then the programs that make the zip code \t{98109} to be a constant instead of \t{AnyStr} are not desirable. On the other hand, we want strings such as \t{today} and \t{weather} to be constants. We overcome both of these issues by devising an output-constrained ranking scheme.


\figref{search} shows our approach where we search for a consistent program in tandem with finding the best program. The algorithm takes as input a DAG $d$,  a list of examples $E$, and a list of unseen inputs $U$, and the outcome is the best program in the DAG that is  consistent with the given examples  (if such a program exists).
The algorithm is a modification to Dijkstra's shortest path algorithm.
 The algorithm maintains a ranked list of at-most $\kappa$ \emph{prefix programs} for each node $\nu$ in the DAG where a prefix program is a path in the DAG from the start node to $\nu$ and the rank of a prefix program is the sum of ranks of all atomic expressions in the path. The atomic expressions are ranked as \t{SubStr} $=$ \t{Replace} $>$ \t{ConstStr} $>$ \t{AnyStr}. Initially, the set of prefix programs for every node is $\emptyset$. The algorithm then traverses the DAG in reverse topological order and for each outgoing edge, it iterates through pairs of prefixes and atomic expressions based on their ranks. For each pair, the algorithm checks if the partial path satisfies the output-uniqueness constraint (\lineref{ddr}) and the generalization constraint (\lineref{idr}). Whenever a consistent pair is found, the concatenation of the atomic expression with the prefix is added to the list of prefixes for the other node in the edge. Finally, the algorithm returns the highest ranked prefix for the target node of the DAG. In the limit $\kappa \rightarrow \infty$, the above algorithm will always find a consistent program if it exists. However, in practice, we found that a smaller value of $\kappa$ is sufficient because of the two pruning steps at \lineref{break1} and \lineref{break2} and because of the ranking that gives least preference to \t{AnyStr} among the other atomic expressions.

\begin{figure}
	\small
	\begin{program}[style=tt, number=true]
		\underline{\t{SearchBestProg}}($\entitydag$, $E$, $U$)
		for\tab each $\nu$ $\in$ $\mathcal{V}(\entitydag)$:
		$\nu$.prefixes = $\emptyset$,
		\untab
		for\tab each $\nu$ $\in$ $\mathcal{V}(\entitydag)$ in reverse topological order:
		for\tab each $e$: $\langle\nu, \nu'\rangle$ $\in$ $\mathcal{E}(\entitydag)$:\linelabel{edge}
		for\tab each $\phi_{\t{sofar}}$ in $\nu$.prefixes.RankedIterator:
		for\tab each $f$ in $\mathcal{W}$($e$).RankedIterator:
		if~\tab Consistent($\phi_{\t{sofar}}$ + $f$, $\nu'$, $E$):\linelabel{ddr}
		if~\tab Generalizes($\phi_{\t{sofar}}$ + $f$, $\nu'$, $U$):\linelabel{idr}
		$\nu'$.prefixes.add($\phi_{\t{sofar}}$ + $f$, ~~$\phi_{\t{sofar}}$.score + $f$.score)
		if $f$ $\neq$ AnyStr: Goto \lineref{break2}\linelabel{break1}
		\untab
		\untab
		\untab
		if AnyStr $\not \in$ $\phi_{\t{sofar}}$: Goto \lineref{edge}\linelabel{break2}
		\untab
		\untab
		\untab
		return $\entitydag$.$\nu_t$.prefixes[0]
	\end{program}
	\caption{Algorithm to find the best consistent program from the DAG learned by \t{LearnURL}.}
	\figlabel{search}
\end{figure}

\begin{theorem}[\textsc{Soundness of GenProg}]\label{thm:1}
	The \t{GenProg} algorithm is sound for all $\lambda_s$, $\lambda_c$ and $\lambda_a$   i.e. given a few input-output examples $\{(i_k,o_k)\}_k$, the learned program $P$ will always satisfy $\forall k.~\llbracket P \rrbracket_{i_k} = o_k$.
\end{theorem}
\textit{Proof sketch:} This holds because the \t{GenDag} algorithm only learns programs that are consistent for each input and \t{Intersect} preserves this consistency across multiple inputs. For learned programs that contain \t{AnyStr} expressions, the \t{SearchBestProg} algorithm ensures soundness because it only adds an atomic expression to a prefix if the combination is consistent with respect to the given examples.

\begin{theorem}[\textsc{Completeness of GenProg}]\label{thm:2}
	The \t{GenProg} algorithm is complete when $\lambda_s$ = True,  $\lambda_c$ = True, $\lambda_a$ = True, and in the limit $\kappa \rightarrow \infty$ where at-most $\kappa$ prefixes are stored for each node in the \t{SearchBestProg} algorithm. In other words, if there exists a program in $L_u$ that is consistent with the given set of input-output examples, then the algorithm will produce one.
\end{theorem}
\textit{Proof sketch:} This is because when $\lambda_s$, $\lambda_c$, and $\lambda_a$ are True, \t{GenDag} will learn all atomic expressions for every edge that satisfy the examples. Since, the DAG structure allows all possible concatenations of atomic expressions, the \t{GenDag} algorithm is complete in this case. \t{Intersect} also preserves completeness, and in the limit $\kappa \rightarrow \infty$, the \t{SearchBestProg} will try all possible paths in the DAG to find a consistent program. Thus, \t{GenProg} does not drop any consistent program and hence, complete.

\begin{theorem}[\textsc{Soundness of LearnUrl}]
The \t{LearnUrl} algorithm	is sound.
\end{theorem}
\textit{Proof sketch: } This is because every layer in the layered search is sound using Theorem \ref{thm:1}.

\begin{theorem}[\textsc{Completeness of LearnUrl}]
The \t{LearnUrl} algorithm is complete in the limit $\kappa \rightarrow \infty$ where at-most $\kappa$ prefixes are stored for each node in the \t{SearchBestProg} algorithm. 
\end{theorem}
\textit{Proof sketch: } This follows because the last layer in the layered search is complete using Theorem \ref{thm:2}.

\newcommand{\pgraph}{predicates graph}
\section{Data Extraction Learning}\seclabel{data-ext}
Once we have a list of URLs, we now need to synthesize a program to extract the relevant data from these web pages. This data extraction is usually done using a query language such as XPath~\cite{xpath} that uses path expressions to select an HTML node (or a list of HTML nodes) from an HTML document. Previous systems such as DataXFormer~\cite{dataxformer,dataxformer1} have considered the absolute XPath obtained from the examples to do the extraction on other unseen inputs. An absolute XPath assumes that all the elements from the root of the HTML document to the target node have that same structure. This assumption is not always valid as web-pages are very dynamic. For instance, consider the \t{weather} data extraction from \exref{weather}. The web pages sometimes have an alert message to indicate events such as storms, and an absolute XPath will fail to extract the weather information from these web pages. More importantly, an absolute XPath will not work for input-dependent data extractions as shown in~\exref{cur}.

Learning an XPath program from a set of examples is called \emph{wrapper induction}, and there exist many techniques~\cite{dalvi09, anton05,stalker,kushmerick97} that solve this problem.  However, none of the previous approaches applied wrapper induction in the context of data-integration tasks that requires generating input-dependent XPath programs. We present a DSL that extends the XPath language with input-dependent constructs. Then, we present an O-PBE based synthesis algorithm that can learn robust and generalizable extraction programs using very few examples. Our synthesis algorithm uses a VSA based technique that allows us to seamlessly integrate with complex string transformation learning algorithms from \secref{url-learn} that are necessary for learning input-dependent extractions. 

Note that it is not always possible to achieve input-dependent data extraction by using a two-phase approach, which first extracts all relevant data from the web-page into a structured format and then, extracts the input-dependent components from this structured data. This is because the data-dependence between the input and the webpage is sometimes hidden, e.g. a stock div element might have id ``msft-price'', which is not directly visible to the users. In these scenarios, it is not possible for the users to identify and provide examples regarding what data should be extracted into the structured format in the intermediate step before the second step of data extraction. Hence, a more integrated approach is required for learning input-dependent extractions. 


\begin{figure*}[t]
	\small
\begin{eqnarray*}
\mbox{Prog } P & := & (\t{name}, \{\pi_1, \cdots, \pi_r\})\\
\mbox{Pred } \pi & := &  \pi_n \; \vert \; \pi_{\t{path}}\\
\mbox{NodePred } \pi_n & := & \pi_a \; \vert \;  \pi_c\\
\mbox{AttrPred } \pi_a & := & [\t{attr(name)} == \h{$\phi$}]\\
\mbox{CountPred } \pi_c & := & [\t{count(axis)} == k]\\
\mbox{PathPred }\pi_{\t{path}} & := & [p]\\
\mbox{Path } p & := &  n_c\;\vert \; n_s \; \vert \; n_a/n_s \; \vert \; p/n_c \\ 
\mbox{Node } n & := & (\t{name}, \t{axis}, \pi_{\t{pos}}, \{\pi_{n_1}, \cdots,  \pi_{n_r}\}) \\
\mbox{PosPred } \pi_{\t{pos}} & := &  [\t{pos} \; (== | \leq )\; k] \; \vert \; \bot \\
\mbox{Axis } \t{axis} & := & \t{Child} \; \vert \; \t{Ancestor} \; \vert \; \t{Left} \; \vert \; \t{Right} \\
\h{\mbox{String } $\phi$} & := &\h{ \t{Same as $\phi$ in \figref{urldslsyntax}}}
\end{eqnarray*}
\caption{ The syntax for the extraction language \lweb, where \t{name} is a string, $k$ is an integer, and $\bot$ is an empty predicate. }
\figlabel{xpathdslsyntax}
\end{figure*}

\begin{figure*}
\small
	\begin{eqnarray*}
		\semw{ (\t{name}, \{\pi_1, \cdots, \pi_r\})} & = & \Filter(\lambda\gamma. \gamma.\t{name} == \t{name} \mbox{ and } \wedge_{k=1}^r \semn{\pi_k} , \t{AllNodes}(\dom))\\
		\semn{[\t{attr}(\t{name}) == \phi]} & = & \entitynodesem.\t{Attr}( \t{name}) == \sem{\phi} \\
		\semn{[\t{count}(\t{axis}) == k]} & = & \t{Len}(\t{Neighbors}(\entitynodesem, \t{axis})) == k \\
		\semn{[p]} & = & \t{Len}(\semnl{p}) > 0 \\
		\semnll{p/n} & = &  \sem{n}(\semnll{p})\\
		\semnll{n} & = &  
		\Filter(\lambda\entitynodesem.\t{Check}_i(n, \entitynodesem), \\
		& & \;\;\;\;\;\; \Flatten(\Map(\lambda\entitynodesem. \t{Neighbors}(\entitynodesem, n.\t{axis}, n.\pi_{\t{pos}})), \nodelist))) \\
		\sem{\phi} & = & \t{Same as in \figref{urldslsemantics}} \\
		\t{Check}_i(n, \entitynodesem) & \equiv & \entitynodesem.\t{name} == n.\t{name} \mbox{ and } \wedge_k\semn{n.\pi_{n_k }}
		\end{eqnarray*}
\caption{The semantics of \lweb, where \t{AllNodes}, \t{Len} and \t{Neighbors} are macros with expected semantics.}
\figlabel{xpathdslsemantics}
\end{figure*}

\subsection{Data Extraction Language \lweb }
\paragraph*{Syntax} \figref{xpathdslsyntax} shows the syntax of \lweb. At the top-level, a program is a tuple containing a name and a list of predicates, where name denotes the HTML tag and
 predicates denote the constraints that the desired ``target'' nodes should satisfy. There are two kinds of predicates---\t{NodePred} and \t{PathPred}. A \t{NodePred} can either be an \t{AttrPred} or a \t{CountPred}. An \t{AttrPred}  has a name and a value. We treat the text inside a node as yet another attribute. The attribute values are not just strings but are string expressions from the DSL $\lurl$ in \figref{urldslsyntax}, which allow the attributes to be computed using string transformations on the input data. This is one of the key differences between the XPath language and \lweb.  A \t{CountPred} indicates the number of neighbors of a node along a particular direction. Predicates can also be \t{PathPred}s denoting existence of a particular path in the HTML document starting from the current node. A path is a sequence of nodes where each node has a name, an axis, a \t{PosPred}, and a list of \t{NodePred}s (possibly empty). The name denotes the HTML tag, axis is the direction of this node from the previous node in the path, and \t{PosPred} denotes the distance between the node and the previous node along the axis. A \t{PosPred} can also be empty ($\bot$) meaning that the node can be at any distance along the axis. In \lweb, we only consider paths that have at-most one node along the \t{Ancestor} axis ($n_a$) and at-most one sibling node along the \t{Left} or the \t{Right} axis ($n_s$). 
 Moreover, the ancestor and sibling nodes can only occur at the beginning of the path.


\paragraph*{Semantics} \figref{xpathdslsemantics} shows the semantics of \lweb. 
A program $P$ is evaluated under a given input data $i$,  on an HTML webpage $\dom$, and it produces a list of HTML nodes that have the same name and satisfy all the predicates in $P$. In this formulation, we use $\entitynodesem$ to represent an HTML node, and it is not to be confused with the node $n$ in the DSL. A predicate is evaluated on an HTML node and results in a Boolean value. Evaluating an \t{AttrPred}
 checks whether the value of the attribute in the HTML node matches the evaluation of the string expression under the given input $i$. A \t{CountPred} verifies that the number of children of the HTML node along the axis (obtained using the \t{Neighbors} macro) matches the count $k$. A \t{PathPred} first evaluates the \t{path} which results in a list of HTML nodes and checks that the list is not empty. 
A path is evaluated step-by-step for each node, where each step evaluation is based on the set of  HTML nodes obtained from the previous steps. Based on the axis of the node in the current step evaluation, the set of HTML nodes is expanded to include all their neighbors along that axis and at a position as specified by the \t{PosPred}. Next, this set is filtered according to the name of the node and its node predicates (using the \t{Check} macro).

\paragraph*{Example} A possible program for the currency exchange rate extraction in \exref{cur} is
	\t{(td, [(td,\\Left,[pos == 1])/}
		 \t{(text,Child},\t{[attr("text") == }$\langle$\textbf{Transformed Date}$\rangle$\t{]))])}.
		  This expression denotes the extraction of an HTML node ($\entitynodesem_1$) with a \t{td}  tag. The path predicate states that there should be another \t{td} HTML node ($\entitynodesem_2$) to the left of $\entitynodesem_1$ at a distance of 1 and it should have a \t{text} child with its \t{text} attribute equal to the transformed date that is learned from the input.

\paragraph*{Design choices} This DSL is only a subset of the XPath language that has been chosen so that it can handle a wide variety of data extraction tasks and at the same time, enables an efficient synthesis algorithm. For example, our top-level program is a single node whereas the XPath language would support arbitrary path expressions. Moreover, paths in XPath do not have to satisfy the ordering criteria that \lweb~enforces and in addition, the individual nodes in the path expression in \lweb~cannot have recursive path predicates. However, we found that most of these constraints can be expressed as additional path predicates in the top-level program and hence, does not restrict the expressiveness.


\newcommand{\entityexamples}{E}
\newcommand{\entityunseen}{U}
\newcommand{\TestNAdd}{\t{TestNAdd}}

\subsection{Synthesis Algorithm}
We now describe the synthesis algorithm for learning a program in \lweb~from examples. 
 Here, the list of input-output examples is denoted as $\entityexamples = \{(\entityii_1, \entitypage_1,\entityhnode_1), (\entityii_2,\entitypage_2,\entityhnode_2), \cdots (\entityii_m,\entitypage_m, \entityhnode_m)\}$, where each example is a tuple of an input $\entityii$, a web page $\entitypage$, and a target HTML node $\entityhnode$, and the list of pairs of unseen inputs and web pages  is denoted as $\entityunseen = \{(\entityii_{m+1}, \entitypage_{m+1}),(\entityii_{m+2}, \entitypage_{m+2}), \cdots,$ $(\entityii_{m+n}, \entitypage_{m+n})\}$.
In this case, the O-PBE synthesis task can be framed as a search problem to find the right set of predicates ($\{\pi_1,\pi_2,\cdots,\pi_r\}$) that can sufficiently constrain the given target example nodes. 

At a high-level, the synthesis algorithm has three key steps: First, it uses the first example to learn all possible predicates for the target node in that example. 
Then, the remaining examples are used to refine these predicates.
Finally, the algorithm searches for a subset of these predicates that uniquely satisfies the given examples and also generalizes well to the unseen inputs.

\begin{figure}
	\small
\begin{eqnarray*}
	\t{LearnTarget}(\gamma) & = & \t{LearnAnchor}(\gamma); \t{LearnChildren}(\gamma); \t{LearnSiblings}(\gamma); \t{LearnAncestors}(\gamma)\\
	\t{LearnChildren}(\gamma) & = & \forall \gamma'~\t{in}~\t{Neighbors}(\gamma, \t{Child}).~ \t{LearnAnchor}(\gamma'); \t{LearnChildren}(\gamma')\\
	\t{LearnSiblings}(\gamma) & = & \forall \gamma'~\t{in}~\t{Neighbors}(\gamma, \{\t{Left}, \t{Right}\}).~ \t{LearnAnchor}(\gamma'); \t{LearnChildren}(\gamma')\\
	\t{LearnAncestors}(\gamma) & = & \forall \gamma'~\t{in}~\t{Neighbors}(\gamma, \t{Ancestor}).~ \t{LearnAnchor}(\gamma'); \t{LearnSiblings}(\gamma')\\
	\t{LearnAnchor}(\gamma) & = &  \t{Anchor}(\gamma.\t{name}, \t{LearnPreds}(\gamma))\\
	\t{LearnAttrPred}(\gamma) & = & \forall \t{attr} ~\t{in}~ \gamma.~\t{AttrPred}(\t{attr}.\t{name}, \t{GenDag}(\t{attr}.\t{value}))\\
	\t{LearnCountPred}(\gamma) & = &  \forall \t{dir}.~\t{CountPred}(\t{dir},\t{Len}(\t{Neighbors}(\gamma, \t{dir})))
\end{eqnarray*}
\caption{Algorithm to transform an HTML document into a \pgraph.}
\figlabel{alg-graph}
\end{figure}

\subsubsection{Learning all predicates} 
For any given example HTML node, there are numerous predicates (especially path predicates) in the language that constrain the target node. In order to learn and operate on  these predicates efficiently, we use a graph data structure called \emph{\pgraph{}} to compactly represent the set of all predicates. This data structure is inspired by the tree data structure used in~\citet{anton05}, but it is adapted to our DSL and algorithm.


Similar to ~\citet{anton05}, to avoid confusion with the nodes in the DSL, we use the term $\Anchor$ to refer to nodes  and the term $\Hop$ to refer to edges in this graph. Hence, a \pgraph{} is a tuple $(A, H, T)$ where $A$ is the list of anchors, $H$ is the list of hops and $T \in A$ is the target anchor. An anchor is a tuple $(n, P)$ where $n$ is the name of the anchor and $P$ is a list of node predicates in the \lweb~language. An edge is a tuple $(a_1, a_2, x, d)$ where $a_1$ is the start anchor, $a_2$ is the end anchor, $x$ is the axis and $d$ is a predicate on the distance between $a_1$ and $a_2$ measured in terms of number of hops in the original HTML document.

\figref{alg-graph} shows the algorithm for transforming an HTML document into a \pgraph{}. We will explain this algorithm based on an example shown in \figref{ex-graph} where the input HTML document is shown on the left, and the corresponding \pgraph{} is shown on the right; the target node is the text node ($T_3$) shown in red. First, it is important to note the difference between these two representations. Although both the HTML document and the \pgraph{} have a tree like structure, the latter is more centered around the target anchor. In the \pgraph{}, all anchors are connected to the target anchor using a minimum number of intermediate anchors that is allowed by the DSL. 
The algorithm first creates an anchor for the target and then learns the anchors for its children, siblings, and ancestors recursively. Learning a child or a sibling will also learn its children in a recursive manner, whereas learning an ancestor will only learn its siblings recursively. Finally, when creating an anchor for an HTML node, all the node predicates of the HTML node are inherited by the anchor, but if there are any attribute predicates, their string values are first converted to DAGs using the $\GenDag$ method in \secref{url}.

After the above transformation, a path $p = a_1/a_2/\cdots a_r$ in the \pgraph{} (where $a_1$ is the target node) represents many different predicates in \lweb~corresponding to different combinations of the node predicates in each anchor $a_k$. We use $\t{predicates}(p)$ to denote the set of all such predicates, e.g. for the path from the target ($T_3$) to the text node $T_2$ in \figref{ex-graph}, $\t{predicates}(p) = \{\pi_{p_1}, \pi_{p_2}, \pi_{p_3} \cdots \}$ where:\\
{\small $\pi_{p_1} = $ \t{[(p, Ancestor, [pos == 1])/(div, Left, [pos == 1])/ (text, Child, [attr("text") = dag$_2$.BestProg])]}\\
 $\pi_{p_2} = $ \t{[(p, Ancestor, [pos == 1])/(div, Left)/(text, Child, [attr("text") = dag$_2$.BestProg])]}\\
 $\pi_{p_3} = $ \t{[(p, Ancestor)/(div, Left)/(text, Child)]}}

  We can define a  partial ordering among predicates generated by a path $p$ as follows: $\pi_{1} \sqsubseteq \pi_{2} $ if the set of all node predicates in  $\pi_{1}$  is a subset of the set of all node predicates in $\pi_{2}$. In the above example, we have $\pi_{p_3} \sqsubseteq \pi_{p_2} \sqsubseteq \pi_{p_1}$.

\begin{definition}[Minimal path predicate]
	Given a path $p$ in the \pgraph{}, a minimal path predicate is the predicate $\pi_p$ encoded by this path such that  $\not\exists \pi_{p'}.~  \pi_{p'} \sqsubseteq \pi_{p}$.
\end{definition}

\begin{definition}[Maximal path predicate]
	Given a path $p$ in the \pgraph{}, a maximal path predicate is the predicate $\pi_p$ such that $\not\exists \pi_{p'}.~ \pi_{p}  \sqsubseteq \pi_{p'}$
\end{definition}
In other words, a \emph{minimal path predicate} is the one that does not have any node predicates for any anchor in the path and a \emph{maximal path predicate} is the one that has all node predicates for every anchor in the path.
For the above example, $\pi_{p_3}$ is the \emph{minimal path predicate} and $\pi_{p_1}$ is the \emph{maximal path predicate} assuming the nodes do not have any other predicates.

\begin{lemma}\label{lem:3}
	Any predicate expressed by the predicates graph, $\Pi$, for an example $(i,w,\gamma)$ will satisfy the example i.e. $\forall \t{path } p \in \Pi.~\forall \pi \in \t{predicates}(p).~\semn{\pi} = \t{True}$.
\end{lemma}
\textit{Proof Sketch: } This lemma is true because \t{LearnTarget} constructs the predicates graph by only adding those nodes and predicates that are in the original HTML document.

\begin{lemma}\label{lem:4}
	The predicates graph, $\Pi$, for an example $(i,w,\gamma)$ can express all predicates in \lweb{} that satisfy the example i.e. $\forall \pi \in L_w.~\semn{\pi} = \t{True} \implies \exists \t{path } p \in \Pi.~\pi \in \t{predicates}(p)$. 
\end{lemma}

\textit{Proof Sketch: } This lemma is true because the traversal of nodes when constructing the predicates graph covers all possible paths in the HTML document that can be expressed in \lweb{}. 

For the implementation, we only construct a portion of the predicates graph that captures nodes that are within a distance $r = 5$ from the target node.

\begin{figure}
\centering
\includegraphics[scale=0.25]{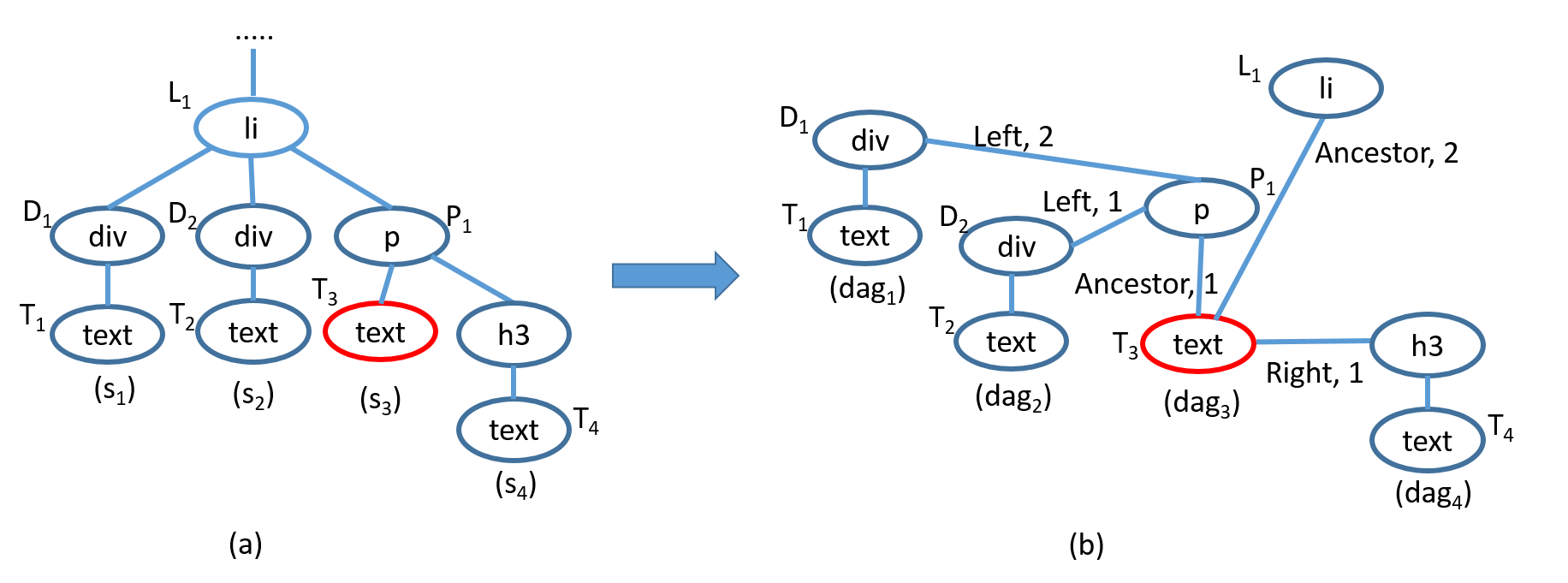}
\caption{An example demonstrating the transformation from an HTML document to a \pgraph{}.}
\figlabel{ex-graph}
\end{figure}

\begin{figure*}
\small
	\begin{eqnarray*}
		\t{IntersectPath}(p,q) & = & \{a_k\}_k ~\t{where}~ a_k = \t{IntersectAnchor}(a_{p_k}, a_{q_k})\\
		\t{IntersectAnchor}(a_p, a_q) & = & \t{Anchor}(a_p.\t{name}, \{\pi_k\}_k) ~\t{where}~ \pi_k = \t{IntersectPred}(\pi_{p_k}, \pi_{q_k})\\ 
		\t{IntersectAttrPred}(\pi_p, \pi_q) & = & \t{AttrPred}(\pi_p.\t{name}, \t{Intersect}(\pi_p.\t{DAG}, \pi_q.\t{DAG}))~ \t{if}~ \pi_p.\t{name} = \pi_q.\t{name}\\
		\t{IntersectPosPred}(\pi_p, \pi_q) & = & \t{PosPred}(\t{==}, \pi_p.k) ~\t{if}~\pi_p.k = \pi_q.k ~\t{else} ~\t{PosPred}(\leq, \t{max}(\pi_p.k, \pi_q.k))\\
		\t{IntersectCountPred}(\pi_p, \pi_q) & = & \t{CountPred}(\pi_p.k) ~\t{if}~\pi_p.k = \pi_q.k
	\end{eqnarray*}
	\caption{Algorithm to intersect two paths in two \pgraph{}s.}
	\figlabel{alg-intersect}
\end{figure*}
\subsubsection{Handling multiple examples}
We, now, have a list of all predicates that constrain the target HTML node for the first example. However, not all of these predicates will satisfy the other examples provided by the user. A simple strategy to prune the unsatisfiable predicates is to create a  \pgraph{} for each example and perform a full intersection of all these graphs. However, this operation is very expensive and has a complexity of $N^m$ where $N$ is the number of nodes in the HTML document and $m$ is the number of examples. Moreover, in the next subsection, we will see that the algorithm for searching the right set of predicates (\figref{ext-min}) will try to add these predicates one-by-one based on a ranking scheme and stops after a satisfying set is obtained. Therefore, our strategy is to refine predicates in the \pgraph{} in a lazy fashion for one path at a time (rather than the whole graph) when the path is required by the \t{SearchBestProg} algorithm.


 We will  motivate this refinement algorithm using the example from \figref{ex-graph}. Suppose that the first example $E_1$ in this scenario has $\entityii_1$ = ``10/16/16'' and $w_1$ as shown in \figref{ex-graph}(a) with $s_{1_1} = $ ``10-16-2016'' and $s_{2_1} = $ ``foo''. As described earlier, one of the possible predicates for this example is: 
 {\small $\pi = $ \t{[(p,Ancestor,[pos==1])/(div,Left,[pos==1])/ (text,Child, [attr("text")=dag$_2$.BestProg])]}.}\\
  Now, suppose that the best program of \t{dag}$_2$ extracts the date (16) in the text of $s_{1_1}$ from the year (16) (rather than  the date 16) in the input $\entityii_1$. Also, assume that the  second example $E_2$ has $\entityii_2= $ ``10/15/16''and $w_2$  similar to $w_1$ but with $s_{1_2} = $ ``bar"  and $s_{2_2} = $ ``10-15-2016''. Clearly, the predicate $\pi$ will not satisfy this new example and hence, we need to refine the path for $\pi$ using the other examples. 
The path for $\pi$, $p = T_{3_1}/P_{1_1}/D_{2_1}/T_{2_1}$, is shown in \figref{ex-path}(a). The algorithm for refining the path is done iteratively  for  one example at a time. For any new example $E_k$, the algorithm will first check if the maximal path predicate of $p$ satisfies the  new example. If it does, then there is no need to refine this path. Otherwise, the algorithm gets all paths in the \pgraph{} corresponding to $E_k$ that satisfies the minimal path predicate of $p$. For the example $E_2$, there are two such paths as shown in \figref{ex-path}(b). The algorithm will then intersect $p$ with each of these paths.
  
  \figref{alg-intersect} shows the algorithm for intersecting two paths. The algorithm  goes through all the anchors in the two paths and intersects each of their node predicates. Note that, since the two paths have the same minimal path predicate, the anchors in the two paths will have the  same sequence of tags.  For intersecting  attribute predicates, the algorithm will intersect their respective DAGs corresponding to the values if their attributes have the same name. For intersecting  position predicates, the algorithm will take the maximum value of $k$ and update the operation accordingly. For intersecting count predicates, the algorithm will return the same predicate if the counts ($k$) are the same. For the example in \figref{ex-path},
   \figref{ex-path}(c)-(d) are the two new resulting paths after the intersection, whose predicates are:\\
   {
   $\pi_1' = $ \t{[(p,Ancestor,[pos==1])/(div,Left,[pos==1])/(text,Child, [attr("text")=(dag$_{2_1}$ $\wedge$ dag$_{2_2}$).TopProg])]}\\
   $\pi_2' = $ \t{[(p,Ancestor,[pos==1])/(div,Left,[pos$\leq$ 2])/(text,Child, [attr("text")=(dag$_{2_1}$ $\wedge$ dag$_{1_2}$).TopProg])]}}

\begin{figure*}
\includegraphics[width=\textwidth]{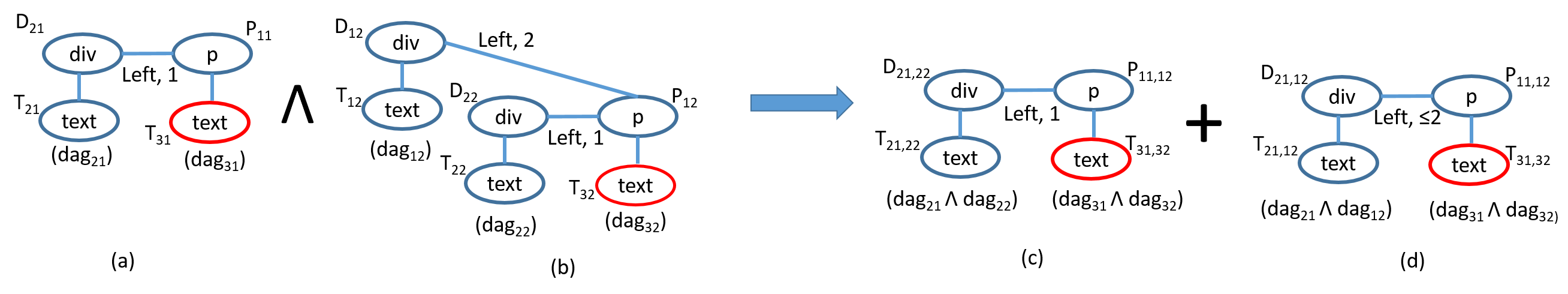}
\caption{Example demonstrating refining a path in \pgraph{} using other examples.}
\figlabel{ex-path}
\end{figure*}

The worst case complexity of this lazy refinement strategy is $N^m$, but in practice, it works well because the algorithm usually accesses few paths in the \pgraph{} because of the ranking scheme and moreover, the path intersection is only required if the original path cannot satisfy the new examples. 

The following lemmas regarding the intersection algorithm hold true by construction. Let $p, q$ be two paths  in the predicates graphs of two different examples $\{(i_1, w_1, \gamma_1), (i_2, w_2, \gamma_2)\}$ that have the same minimal path predicate, and  let $r$ be the path obtained after the intersection. 

\begin{lemma}\label{lem:5}
	Any predicate obtained after intersecting two paths satisfies all the examples. Formally, $\forall \pi \in \t{predicates}(r).~\llbracket \pi \rrbracket_{i_1}(\gamma_1) = \t{True} \wedge \llbracket \pi \rrbracket_{i_2}(\gamma_2) = \t{True}$.
\end{lemma}

\begin{lemma}\label{lem:6}
	Intersecting two paths preserves all predicates that satisfy the examples. Formally, $\forall \pi.~\pi \in \t{predicates}(p) \wedge \pi \in \t{predicates}(q) \implies \pi \in \t{predicates}(r)$.
\end{lemma}


%
\begin{figure}
	\small
	\begin{program}[style=tt,number=true]
		\stab{1mm}\underline{$\Search$}($\entityallpreds$, $\entityexamples$, $\entityunseen$) \nonumber
		$\entityallpreds'$ = $\emptysetC$; $\RankPredicates$($\entityallpreds$)
		for\tab each $\entitypred$ in $\entityallpreds$.$\SortedIterator$:
		$\entitypathslist$ = $\Refine$(Path($\entitypred$), $\entityexamples$)\linelabel{xpath-refine}
		for\tab each $\entitypred'$ in $\entitypathslist$
		done = $\TestNAdd$($\entitypred'$,  $\entityallpreds'$,  $\entityexamples$, $\entityunseen$) 
		if  (done): return ($\entityexamples$[0].$\entityhnode$.Name, $\entityallpreds'$)\linelabel{xpath-min-end}
		\untab
		\untab
	\end{program}
%
	\caption{The algorithm for computing the best predicate set using output-constrained ranking.}
	\figlabel{ext-min}
\end{figure}

\subsubsection{Output-constrained ranking of predicates}
We now have a list of predicates that satisfy the first example and we have an algorithm to refine predicates based on the other examples provided by the user. However, only some of these predicates are desirable as some predicates might not generalize to unseen inputs and some others might not be required to constrain the target nodes. 
\figref{ext-min} shows the algorithm  that uses \emph{output-constrained ranking} to find a subset of the predicates that generalizes well. The algorithm takes as inputs a set of predicates $\Pi$, a list of input-output examples $E$, and a list of unseen inputs $U$. It uses the unseen inputs as a test set to prune away predicates that do not generalize well. The algorithm iterates through the list of all predicates based on a ranking scheme and adds the predicate if there is at-least one node in each test document that satisfies the predicates added so far.
This process is stopped if we find a set of predicates that uniquely constrains the target nodes in the provided examples.

\paragraph*{Ranking Scheme} 
We use the following criterion that gives higher priority to  predicates that best constrain the set of target nodes and at the same time, capture user intentions: (i) Attribute predicates with more sub-string expressions on input data are preferred over other attribute predicates, (ii) Left siblings nodes are preferred over right siblings because usually websites contain descriptor information to the left of values and in most cases, these descriptor information tend to be the same across websites, and (iii) Nodes closer to target are preferred over farther nodes.



\begin{theorem}[\textsc{Soundness}]
The data extraction synthesis algorithm is sound i.e. given some input-output examples $\{(i_k,w_k,\gamma_k)\}_k$, the program $P$ that is learned by the algorithm will always satisfy $\forall k.~\llbracket P\rrbracket_{i_k}(w_k) = \{\gamma_k\}$.
\end{theorem}
\textit{Proof Sketch: } This theorem follows directly from Lemmas \ref{lem:3} and \ref{lem:5}.

\begin{theorem}[\textsc{Completeness}]
The data extraction synthesis algorithm is complete i.e. if there exists a program in $L_w$ that satisfies the given I/O examples, then the algorithm is guaranteed to find one program that satisfies the examples.
\end{theorem}

\textit{Proof Sketch: } This holds because the predicates graphs  created for the examples are complete using Lemma \ref{lem:4}. And the algorithm for 
refining a path intersects the path with all similar paths (that satisfy the minimal path predicate) in the other examples and \t{IntersectPath} preserves all predicates that can be satisfied by the examples (Lemma  \ref{lem:6}).

Note, that the $\Search$ algorithm cannot influence the soundness or completeness argument because the set of all predicates obtained after refining every path in the predicates graph is a sound and complete solution. $\Search$ only influences how well it generalizes to unseen inputs. 

\paragraph*{A Note on adding new rows and changing data sources} When adding new rows to the spreadsheet after the integration task has been performed, there might be concerns about whether the joined data for the old rows will be obsolete.  For example, in \exref{stock} or \exref{weather}, the stock values or the weather information would change presumable every second. However, the user still does not need to re-provide updated examples. This is because \sys{} learns programs in the DSLs and it can just re-execute the learned program directly and compute results for the new rows. Only in cases if the website DOM structure changes (which does not happen too often for major websites), the user would need to re-update the previously provided examples.
\section{Evaluation}\seclabel{eval} 
In this section, we evaluate $\sys$ on a variety of real-world web integration tasks. In particular, we evaluate whether the DSLs for URL learning and data extraction are expressive enough to encode these tasks, the efficiency of the synthesis algorithms, the number of examples needed to learn the programs, and the impact of layered  version spaces and output-constrained ranking. 
%

\paragraph*{Benchmarks}
We collected 88 web-integration tasks taken from online help forums and the Excel product team. These tasks cover a wide range of domains including finance, sports, weather, travel, academics, geography, and entertainment.
 Each benchmark has 5 to 32 rows of input data in the spreadsheet. These 88 tasks decompose into 62 URL learning tasks and 88 extraction tasks. For some benchmarks, we had scenarios with examples and webpages provided by the Excel team. For other benchmarks, we chose alternate sources for data extraction and provided examples manually, but they were independent of the underlying learning system. The set of benchmarks with unique URLs is shown in \figref{bench1}. 

\begin{figure*}
\scriptsize
\begin{tabular}{|l | l | l | l|l|}
\hline
{\#} & {\bf Description} & {\bf \#R} & {\bf Example data item(s)} & {\bf Website Domain}\\
\hline
{1-2}& {ATP players to ages/countries} & {5} & {(age 29) | Serbia and Montenegro} & {wikipedia} \\
{3-4}& {ATP players to latest tweet/total tweets} & {5} & {Long tweet message | 2,324} & {twitter} \\
{5-7}& {ATP players to single titles/ranking/W-L} & {20} & {7 | ATP Rank \#1 | 742-152} & {espn}\\
{8}& {ATP players to number of singles based on year} & {7} & {7} & {espn.go} \\
{9}& {ATP players to rankings} & {5} & {1} & {atpworldtour} \\
{10}& {ATP players to rankings} & {5} & {\#1} & {tennis}\\
{11}& {ATP players to career titles} & {5} & {66 (7th in the Open Era)} & {wikipedia}\\
{12}& {Addresses to population} & {7} & {668,342 (100\% urban, 0\% rural).} & {city-data} \\
{13-14}& {Addresses to population/zipcode} & {7} & {786,130 | 98101 Zip Code} &{zipcode}\\
{15-16}& {Addresses to weather/weather based on date} & {7} & {51$^\circ$ | 51$^\circ$} & {accuweather}\\
{17}& {Addresses to weather} & {7} & {57$^\circ$F} & {timeanddate} \\
{18}& {Addresses to weather} & {7} & {52} & {weather}\\
{19-23}& {Addresses to weather stats} & {6} & {49|Partly sunny|74$^\circ$|62|68$^\circ$} & {accuweather} \\
{24}& {Airport code to airport name} & {5} & {Boston Logan International Airport (BOS)} & {virginamerica} \\
{25-26}& {Airport code to delay status/name} & {5} & {Very few delays | Logan International} & {flightstats} \\
{27}& {Airport code to terminal information} & {5} & {Terminal B - Gates B4 - B19} & {aa}\\
{28}& {Albums to genre} & {5} & {Rock} & {wikipedia}\\
{29-31}& {Authors to paper citations, max citation, title} & {5} & {Cited by 175 | Cited by 427 | Syntax-guided }  & {scholar.google}\\
{32-35}& {Authors to different data from DBLP} & {5} & {Demo...|Alvin Cheung(12)|PLDI(9)|POPL(2)} & {dblp} \\
{36}& {Cities to population} & {6} & {21,357,000} & {worldpopulation}\\
{37}& {Company symbols to 1 year target prices} & {6} & {65}  & {nasdaq}\\
{38}& {Company symbols to stock prices} & {6} & {59.87} & {finance.yahoo} \\
{39}& {Company symbols to stock prices} & {6} & {59.87} & {money.cnn} \\
{40}& {Company symbols to stock prices} & {6} & {59.87} & {quotes.wsj} \\
{41}& {Company symbols to stock prices} & {6} & {59.87} & {google} \\
{42}& {Company symbols to stock prices} & {6} & {59.00} & {marketwatch}\\
{43}& {Company symbols to stock prices} & {6} & {59.95} & {nasdaq}\\
{44}& {Company symbols to stock prices} & {6} & {\$59.87} & {google}\\
{45}& {Company symbols to stock prices on date} & {6} & {60.61} & {google} \\
{46}& {Company symbols to stock prices on date} & {6} & {60.61} & {yahoo}\\
{47}& {Country names to population} & {6} & {1,326,801,576} & {worldometers}\\
{48}& {Country names to population} & {6} & {1,336,286,256} & {wikipedia}\\
{49}& {Cricket results for teams on different dates} & {5} & {Australia won by 3 wickets} & {espncricinfo}\\
{50-52}& {Cricket stats for two different teams} & {5} & {90 | 24 | 31.67} & {espncricinfo}\\
{53}& {Currency exchange values} & {8} & {66.7870} & {finance.yahoo} \\
{54}& {Currency exchange values} & {8} & {INR/USD = 66.84699} & {moneyconverter}\\
{55}& {Currency exchange values} & {8} & {66.7800} & {bloomberg} \\
{56}& {Currency exchange values} & {8} & {66.8043 INR} & {investing} \\
{57}& {Currency exchange values} & {8} & {66.779} & {xe} \\
{58}& {Currency exchange values} & {8} & {66.7914} & {exchange-rates}\\
{59}& {Currency exchange values based on date} & {8} & {64.1643 INR} & {investing} \\
{60}& {Currency exchange values based on date} & {8} & {66.778}  & {exchange-rates}\\
{61}& {Currency exchange values based on date} & {8} & {64.231} & {investing}\\
{62}& {Flight distance between two cities} & {5} & {2,487 Miles} & {cheapoair} \\
{63-65}& {Flight fares between two cities/cheapest/airline} & {5} & {\$217 | Wednesday | \$237} & {farecompare} \\
{66}& {Flight fares between two cities} & {5} & {\$257} & {cheapflights} \\
{67}& {Flight travel time between two cities} & {5} & {6 hrs 11 mins} & {travelmath} \\
{68}& {Flight travel time between two cities} & {5} & {5 hours, 38 minutes} & {cheapflights}\\
{69}& {NFL teams to rankings} & {5} & {4-5, 3rd in AFC East} & {espn}\\
{70}& {NFL teams to rankings} & {5} & {26th} & {cbssports} \\
{71}& {NFL teams to rankings} & {5} & {26.7} & {nfl}\\
{72}& {NFL teams to rankings} & {5} & {\#5} & {teamrankings}\\
{73}& {NFL teams to stadium names} & {32} & {New Era Field} & {espn}\\
{74}& {Nobel prize for different subjects based on year} & {5} & {Richard F. Heck} & {wikipedia}\\
{75}& {Nobel prize for different years based on subject} & {5} & {Richard F. Heck} & {nobelprize}\\
{76}& {Nobel prize for different years based on subject} & {5} & {Richard F. Heck} & {nobelprize} \\
{77-78}& {Novels to authors/genre} & {7} & {Charles Dickens | Historical novel} & {wikipedia} \\
{79}& {Novels to authors} & {8} & {Charles Dickens} & {goodreads} \\
{80}& {\# of coauthored papers between two authors} & {10} & {Rastislav Bodik (6)} & {dblp}\\
{81}& {Number of daily flights between two cities} & {5} & {26} & {cheapoair}\\
{82}& {People to profession} & {5} & {Principal Researcher} & {zoominfo} \\
{83-84}& {Real estate properties to sale prices} and stats & {5} & {\$1,452 | 1,804 sqft} & {zillow} \\
{85}& {Stock prices with names from same webpage} & {6} & {7.60} & {marketwatch}\\
{86-88}& {Video names to youtube stats} & {5} & {11,473,055 | 2,672,818,718 views | Jul 15, 2012} & {youtube} \\
\hline
\end{tabular}
\caption{The set of web data integration benchmarks with a brief description along with number of input rows (\#R), the example data item(s), and the website domain from which the data is extracted.}
\figlabel{bench1}
\end{figure*}

\paragraph*{Experimental Setup}
We implemented $\sys$ in C\#. All experiments were performed using a dual-core Intel i5 2.40 GHz CPU with 8GB RAM.
For each component, we incrementally provided examples until the system learned a program that satisfied all other inputs, and we report the results of the final iteration. 

\subsection{URL learning}
For each URL learning task, we run   the layered search using 4 different configurations for the layers: 1. \t{L1 to L4}, 2. \t{L2 to L4}, 3. \t{L3 to L4}, and 4. \t{Only L4} where \t{L1}, \t{L2}, \t{L3}, and \t{L4} are as defined in \figref{learnealgo}. The last configuration essentially compares against the VSA algorithm used in FlashFill (except for the new constructs). 

\figref{url}(a) shows the synthesis times\footnote{excluding the time take to run search queries on the search engine.}  required for learning URLs.
We categorize the benchmarks based on the layer that has the program required for performing the transformation. 
We can see that for the \t{L1 to L4} configuration, $\sys$ solves all the tasks in less than 1 second. The \t{L2 to L4} configuration performs equally well whereas the performance of \t{only L4} configuration is much worse. Only 28 benchmarks complete when given a timeout of 2 minutes. Note that none of the URL learning tasks need the L4 configuration in our benchmarks, but we still allow for using L4 for two reasons: i) completeness of the synthesis algorithm for our DSL for other benchmarks in future, and ii) comparison against a no layered approach (a baseline similar to FlashFill like VSA algorithms).

\begin{figure}
\begin{tabular}{c c}
\begin{minipage}{0.45\linewidth}
	\includegraphics[width=0.9\linewidth]{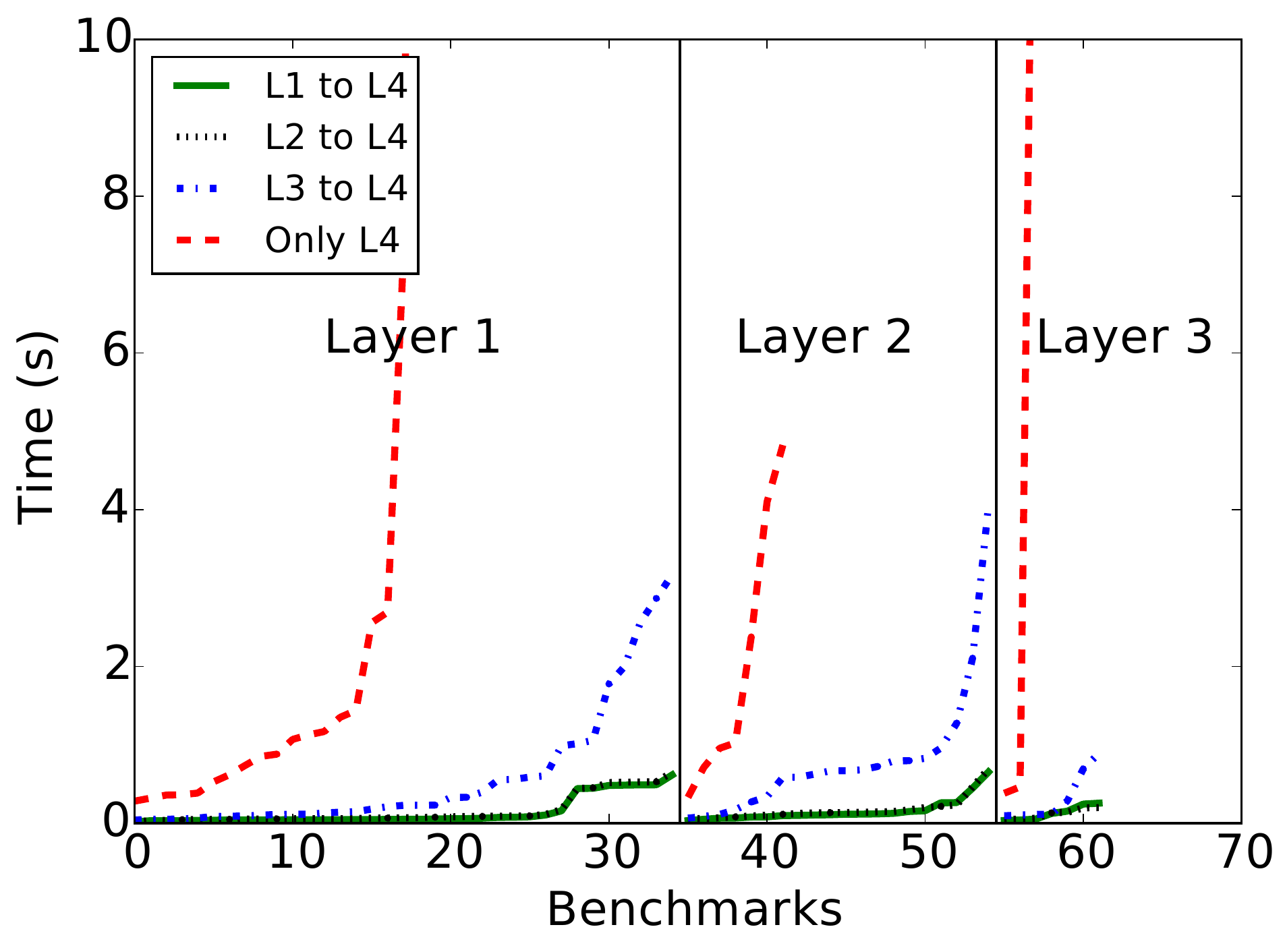}

\end{minipage}
&
\begin{minipage}{0.45\linewidth}
	\includegraphics[width=0.9\linewidth]{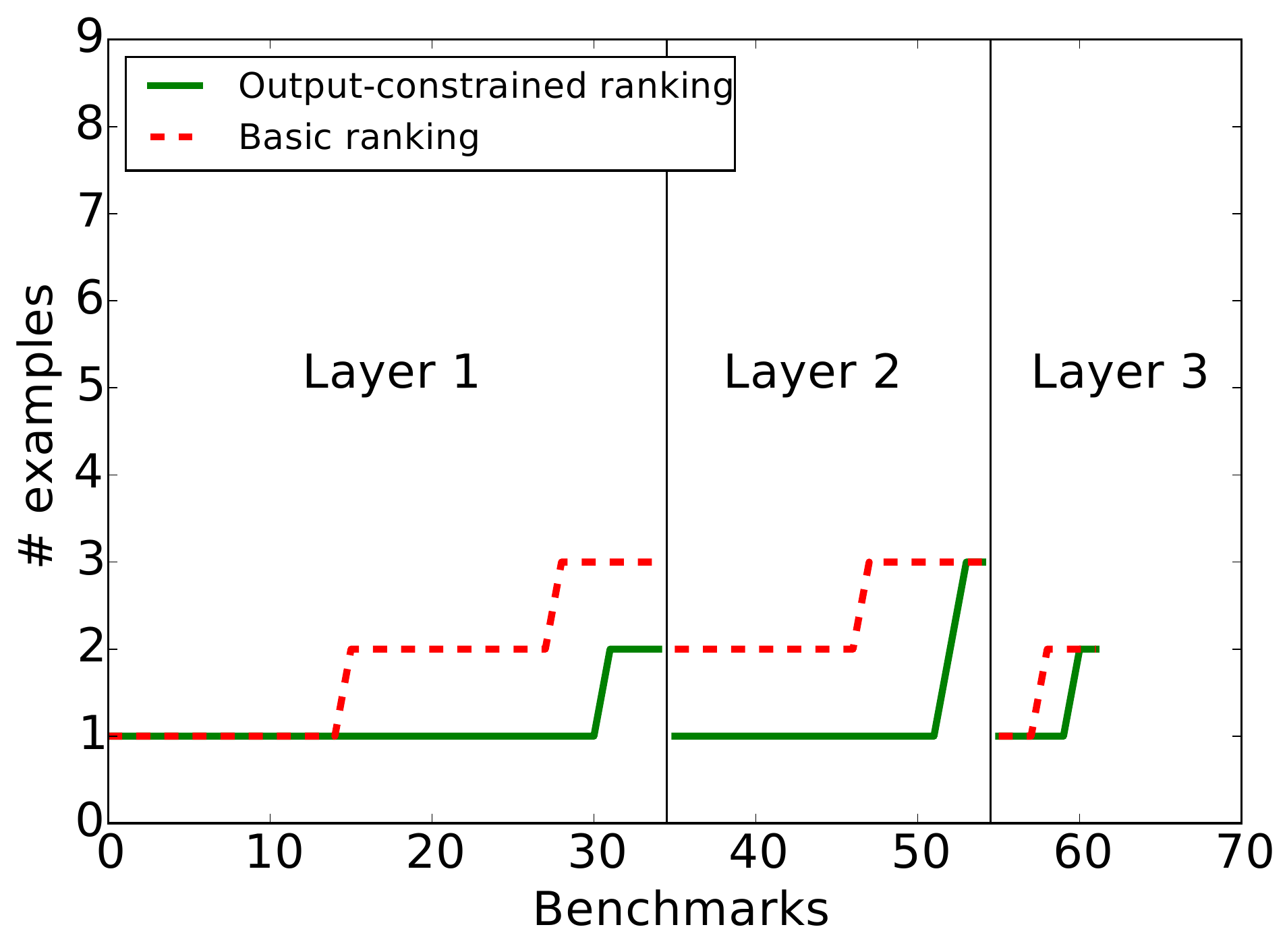}
\end{minipage}
\\
(a) & (b)
\end{tabular}
	\caption{(a) The synthesis times and (b) the number of examples required for learning URL  programs. The benchmarks are categorized based on the layer that contains the program required for performing the transformation.}
	\figlabel{url}
\end{figure}

For these tasks, the length of the URLs is in the range of 23 to 89. Regarding DSL coverage, about 50\% of the benchmarks require \t{AnyStr}, 88\% require \t{SubStr}, 38\% require \t{Replace}, and all benchmarks require \t{ConstStr} expressions.
Thus, all DSL features are needed for different subsets of benchmarks. Moreover, note that the 50\% of the benchmarks that require \t{AnyStr} expressions cannot be learned by traditional PBE techniques and hence, require the O-PBE formulation. 

We also perform an experiment to evaluate the impact on generalization with the  \emph{output-constrained ranking} scheme. For this experiment, we use the \t{L1 to L4} configuration for the layered search and report the results in \figref{url}(b). With output-constrained ranking, 85\% of the benchmarks require only 1 example whereas without it, only 29\% of benchmarks can be synthesized using a single example. Note that the \t{L1 to L4} layered search already has a strong prior that the programs in earlier layers are more desirable, and the output-constrained ranking is able to further reduce the number of examples required.
 

 To evaluate the scalability with respect to the number of examples, we performed an experiment where we took a benchmark that has 32 rows in the spreadsheet and incrementally added more examples. The results are shown in \figref{longex}(a). Although the theoretical complexity of the algorithms is exponential in the number of examples, in practice, we observed that the performance scales almost linearly. We attribute this to the sparseness of the DAGs learned during the layered search. 
 
\begin{figure*}[t]
	\includegraphics[width=0.99\textwidth]{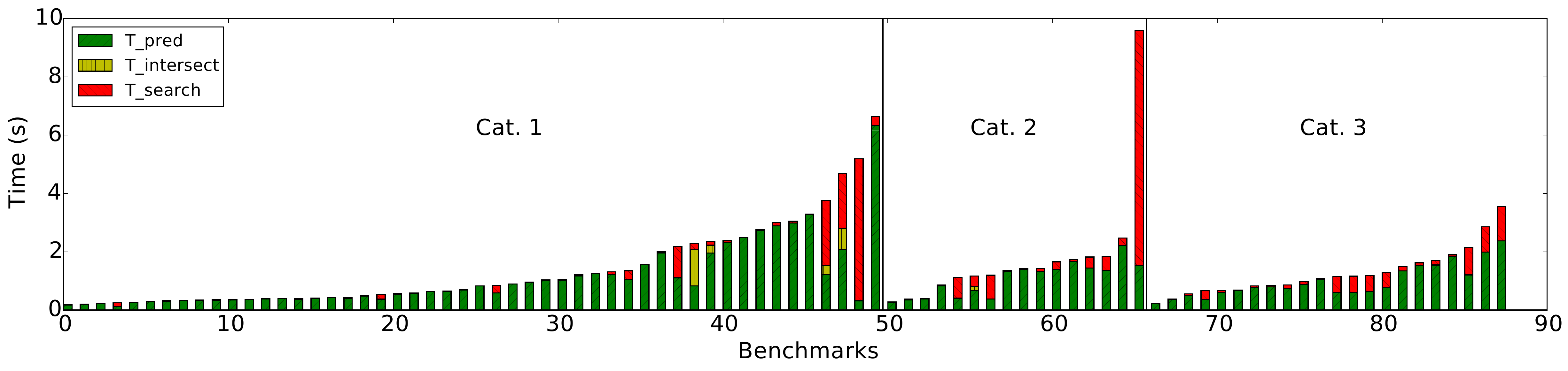}
	\caption{The synthesis times for learning data extraction programs by \sys{} on the 88 benchmarks.}
	\figlabel{xpath}
\end{figure*}

\begin{figure}
	\begin{tabular}{c c}
		\begin{minipage}{0.45\linewidth}
			\includegraphics[width=0.9\linewidth]{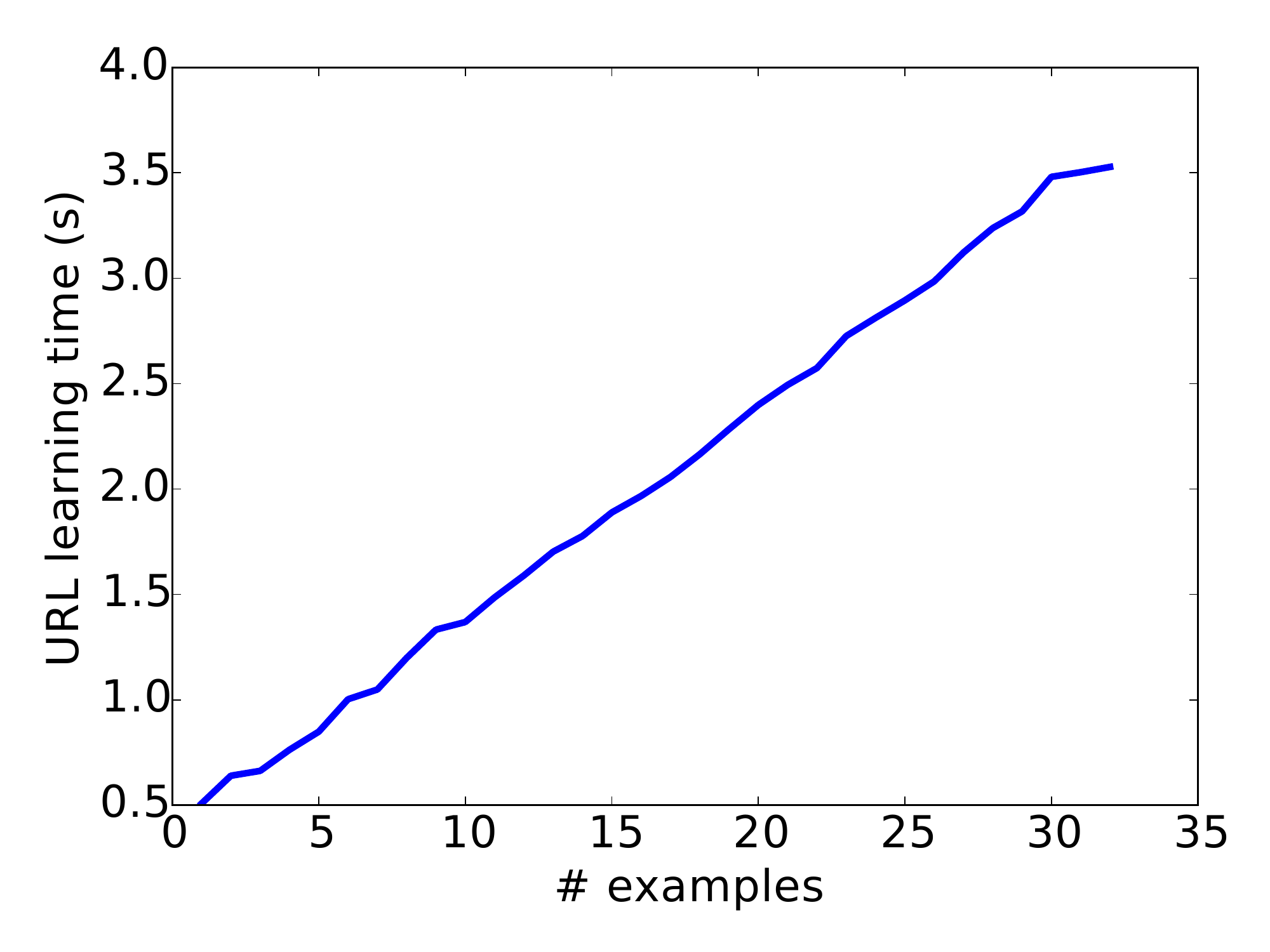}
			
		\end{minipage}
		&
		\begin{minipage}{0.45\linewidth}
			\includegraphics[width=0.9\linewidth]{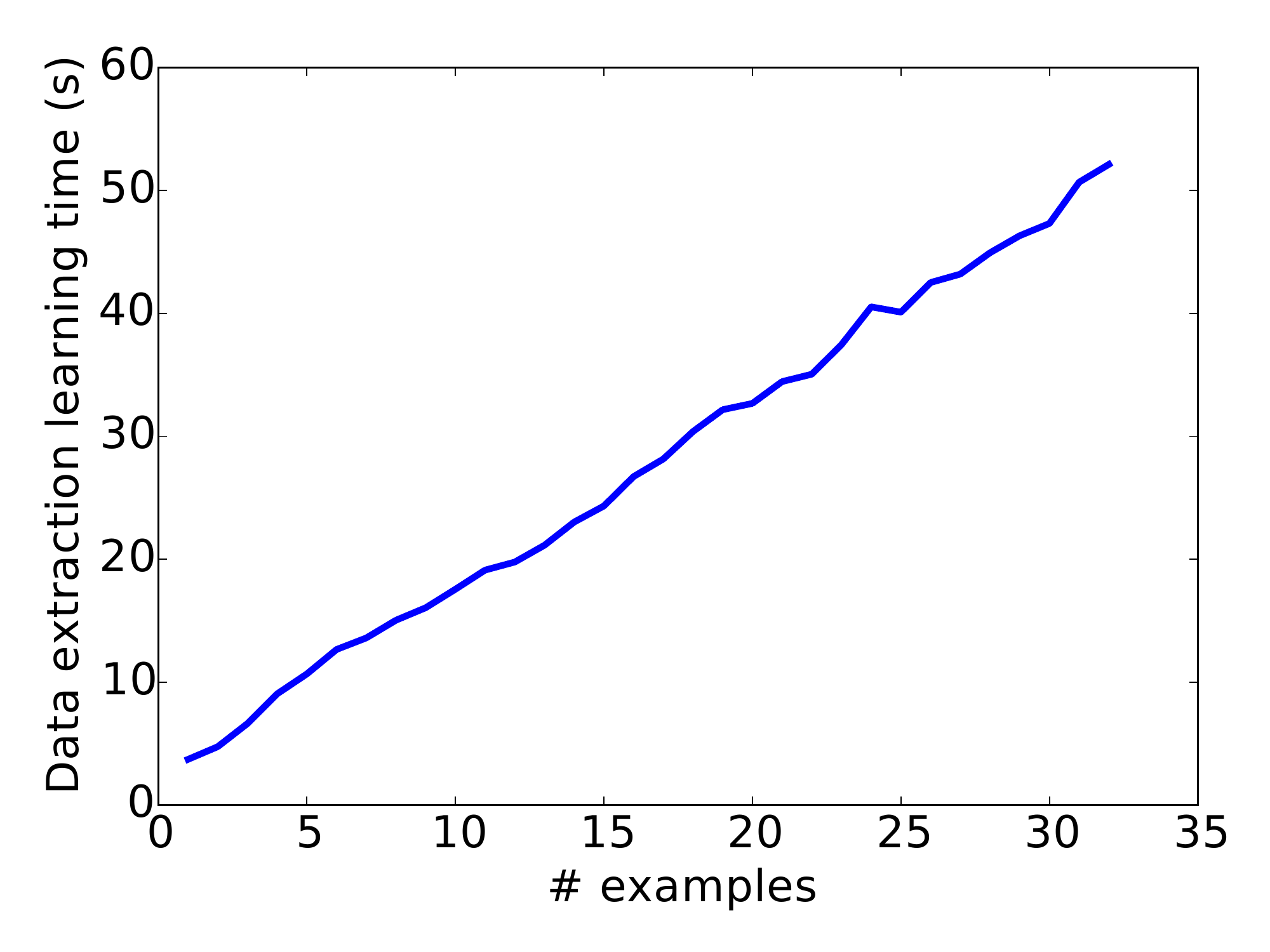}
		\end{minipage}
		\\
		&\\
		(a) & (b)
	\end{tabular}
	\caption{The synthesis time vs the number of examples for URL learning (a) and data extraction learning (b).}
	\figlabel{longex}
\end{figure}

\begin{figure}
	\centering
	\includegraphics[width=0.45\textwidth]{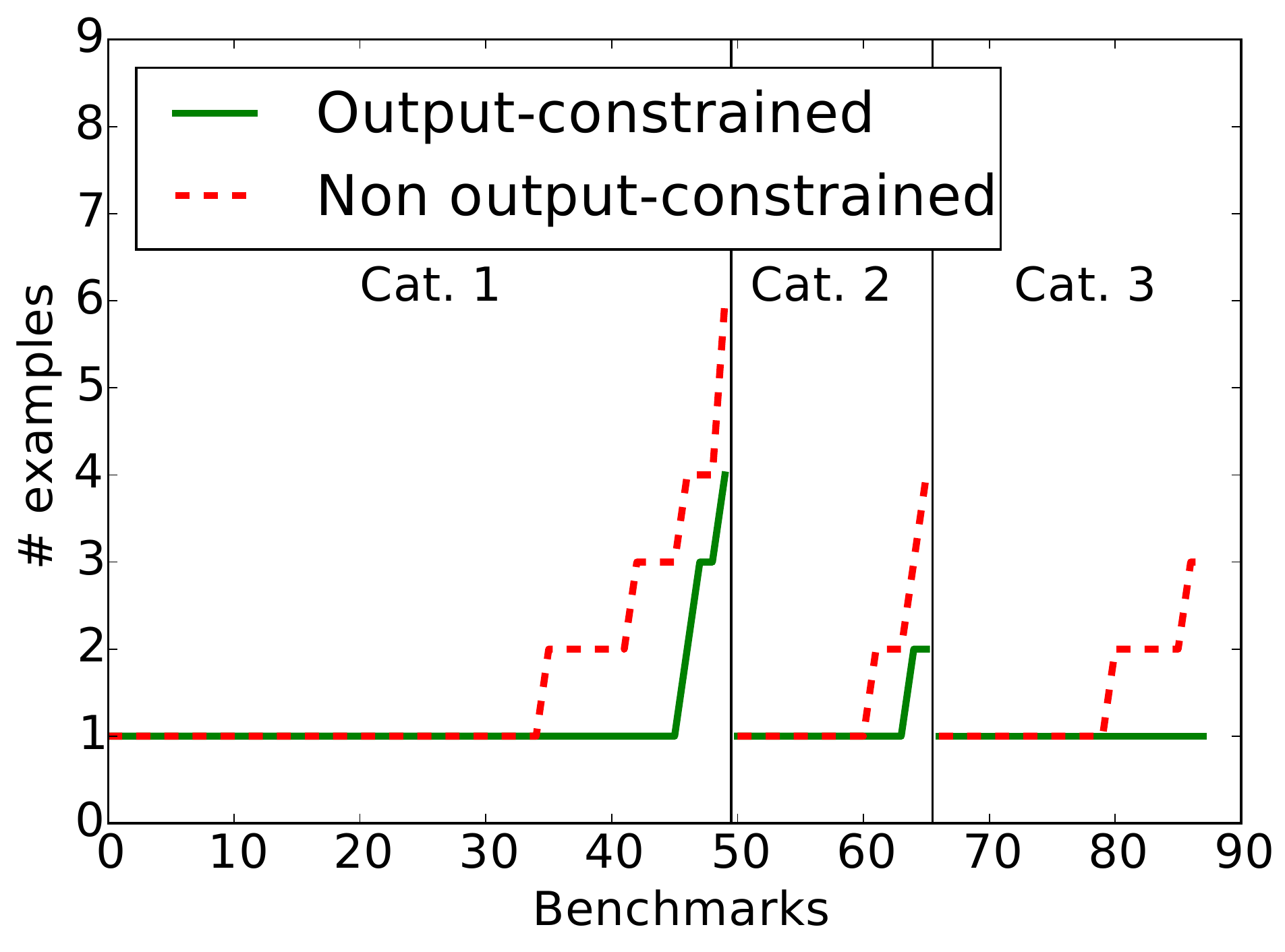}
	\caption{The number of examples needed to learn the data extraction programs.}
	\figlabel{xpathex}
\end{figure}

\subsection{Data Extraction}
We now present the evaluation of our data extraction synthesizer. We categorize the benchmarks into three categories. The first category consists of benchmarks where the data extraction can be learned using an absolute XPath. The second category includes the benchmarks that cannot be learned using an absolute XPath, but by using relative predicates (especially involving nearby strings). The last category handles the cases where the data extraction is input-dependent. 

\figref{xpath} shows the synthesis times\footnote{excluding the time taken to load the webpages} for all 88 benchmarks. The figure also splits the synthesis time into the time taken for learning the predicates graphs (\t{T\_pred}),  for intersecting the predicates graphs if there are multiple examples (\t{T\_intersect}) and finally, for searching  the right set of predicates (\t{T\_search}). $\sys$ can learn the correct extraction for all the benchmarks in all the three categories showing that the DSL is very expressive. Moreover, it can learn them very quickly---97\% of benchmarks take less than 5 seconds. \figref{xpathex} shows the number of examples required and also compares against non output-constrained ranking. It can be seen that with output-constrained ranking, 93\% of benchmarks require only 1 example. 

Most of the synthesis time is actually spent in generating the predicates graphs and this time is proportional to the number of attribute strings in the HTML document since \sys{} will create a DAG for each string. In our examples, the number of HTML nodes we consider for the predicates graph is in the range of 30 to 1200 with 3 to 200 attribute strings. For some benchmarks, the time spent in searching for the right set of predicates dominates. These are the benchmarks that require too many predicates to sufficiently constrain the target nodes. The actual time spent on intersecting the predicates graph is not very significant. For benchmarks that require multiple examples, we found that the time spent on intersection is only 15\% of the total time (on average) due to our lazy intersection algorithm. 

\figref{longex}(b) shows how the performance scales with respect to the number of examples. Similar to URL learning, the performance scales almost linearly as opposed to exponentially. We attribute this to our lazy intersection algorithm.

\section{Related Work}

\noindent
\textbf{\textit{Data Integration from Web:}}
DataXFormer~\cite{dataxformer,dataxformer1} performs semantic data transformations using web tables and forms. Given a set of input-output examples, it searches a large index of web tables and web forms to find a consistent transformation.
For web forms, it performs a web search consisting of input-output examples to identify web forms that might be relevant, and then uses heuristics to invoke the form with the correct set of inputs. 
Instead of being completely automated, $\sys$, on the other hand, allows users to identify the relevant websites and point to the desired data on the website, which allows $\sys$ to perform data integration from more sophisticated websites. Moreover, $\sys$ allows for joining data based on syntactic transformations on inputs, whereas DataXFormer only searches based on exact inputs.

WebCombine~\cite{webcombine} is a PBD web scraping tool for end-users that allows them to first extract logical tables from a webpage, and then provide example interactions for the first table entry on another website. It uses Ringer~\cite{ringer} as the backend record and replay engine to record the interactions performed by the user on a webpage. The recorded interaction is turned into a script that can be replayed programmatically. WebCombine parameterizes the recorded script using 3 rules that parameterize xpaths, strings, and frames with list items. 
Vegemite~\cite{vegemite} uses direct manipulation and PBD to allow end-users to easily create web mashups to collect information from multiple websites. It first uses a demonstration to open a website, copy/paste the data for the first row into the website, and records user interactions using the CoScripter~\cite{coscripter} engine. The learnt script is then executed for remaining rows in the spreadsheet. The XPath learning does not need to be complex since all rows go to the same website and the desired data is at the same location in the DOM.

Instead of recording user demonstrations, $\sys$ uses examples of URLs (or search queries) to learn a program to get to the desired webpages. The demonstrations-based specification has been shown to be challenging for users~\cite{pbdfail} and many websites do not expose such interactions. These systems learn simpler XPath expressions for data extraction, whereas $\sys$ can learn input data-dependent Xpath expressions that are crucial for data integration tasks. Moreover, these systems assume the input data is in a consistent format that can be directly used for interactions, whereas $\sys$ learns additional transformations on the input for both learning the URLs and data-dependent extractions.

\noindent
\textbf{\textit{Programming By Examples (PBE) for string transformations:}} Data Wrangler~\cite{wrangling} uses examples and predictive interaction~\cite{heer2015predictive} to create reusable data transformations such as map, joins,
aggregation, and sorting. There have also been many recent PBE systems such as FlashFill~\cite{popl11}, BlinkFill~\cite{blinkfill}, and FlashExtract~\cite{flashextract} that use VSA~\cite{flashmeta} for efficiently learning string transformations from examples. Unlike these systems that perform data transformation and extraction from a single document, $\sys$ joins data between a spreadsheet and a collection of webpages. $\sys$ builds on top of the substring constructs introduced by these systems to perform both URL learning and data extraction. Moreover, $\sys$ uses layered version spaces and a output-constrained ranking technique to efficiently synthesize programs.

There is another PBE system that learns relational joins between two tables from examples~\cite{vldb12}. It uses a restricted set of relational algebra to allow using VSA for efficient synthesis. Unlike learning joins between two relational sources, $\sys$ learns joins between a relational data source (spreadsheet) and a semi-structured data source (multiple webpages).

\noindent
\textbf{\textit{Wrapper Induction: }} Wrapper induction~\cite{kushmerick97} is a technique to automatically extract relational information from webpages using labeled examples. There exists a large body of research on wrapper induction with some techniques using input-output examples to learn wrappers~\cite{kushmerick97,dalvi09,anton05,stalker,hsu98,flashextract} and others~\cite{webcombine,roadrunner,zhai05} perform unsupervised learning using pattern mining or similarity functions. Some of these techniques~\cite{anton05,dalvi09,flashextract} are based on Xpath similar to \sys{} whereas some techniques such as~\cite{kushmerick97} treat webpages as strings and learn delimiter strings around the desired data from a fixed class of patterns. 
The main difference between any of these techniques and \sys{} is that the extraction language used by $\sys$ is more expressive as it allows richer Xpath expressions that can depend on inputs.

\noindent
\textbf{\textit{Program Synthesis: }} The field of program synthesis has seen a renewed interest in recent years~\cite{sygus,synthnow}. In addition to VSA based approaches, several other approaches including constraint-based~\cite{sketch}, enumerative~\cite{transit}, stochastic~\cite{stochastic}, and finite tree automata based techniques~\cite{fta,afta} have been recently developed to synthesize programs in different domains. Synthesis techniques using examples have also been developed for learning data structure manipulations~\cite{storyboard}, type-directed synthesis for recursive functions over algebraic datatypes~\cite{typesynthesis,typepbe}, transforming tree-structured data~\cite{treesynthesis}, and interactive parser synthesis~\cite{parsersynthesis}.
These approaches are not suited for learning URLs and data extraction programs because the DSL operators such as regular expression based substrings and data-dependent xpath expressions are not readily expressible in these approaches and enumerative approaches do not scale because of large search space. 

\section{Limitations and Future Work}
There are certain situations under which \sys{} may not be able to learn a data integration task. 
First, since all of our synthesis algorithms are sound, \sys{} cannot deal with noise. For example, if  any input data in the spreadsheet is misspelled or has a semantically different format, then \sys{} may not be able to learn such string transformation  programs. Our system can handle syntactic data transformations and partial semantic transformations, but not fully semantic transformations. 
For instance, in \exref{country} if the input country name was ``US'' instead of ``United States'', then \sys{} would not be able to learn such programs. 
As future work, we plan to use recent neural program synthesis techniques~\cite{iclr17,snapl17,robustfill} to learn a noise model based on semantics and web tables~\cite{semajoin}, and incorporate this model into the synthesis algorithm to make it more robust with respect to noise, small web page discrepancies, and different semantic data formats. 
In \sys{}, we assume that the webpages containing the desired data have URLs that can be programmatically learned. However, there are also situations that require complex interactions such as traversing through multiple pages possibly by interacting with the webpage UI before getting to the page that has the desired data. In future, we want to explore integrating the techniques in this paper with techniques from record and replay systems such as Ringer~\cite{ringer} to enable data-dependent replay.  
\section{Conclusion}
We presented $\sys$, a PBE system for joining semi-structured web data with relational data. The key idea in $\sys$ is to decompose the task into two sub-tasks: URL learning and data extraction learning. We frame the URL learning problem in terms of learning syntactic string transformations and filters, whereas we learn data extraction programs in a rich DSL that allows for data-dependent Xpath expressions. The key idea in the synthesis algorithms is to use layered version spaces and output-constrained ranking to efficiently learn the programs using very few input-output examples. We have evaluated $\sys$ successfully on several real-world web data integration tasks.

\begin{acks}
We would like to thank Armando Solar-Lezama, Ben Zorn, and anonymous reviewers for their constructive feedback and insightful comments. We would also like to thank the members of the Microsoft Excel and Power BI teams for their helpful feedback on various versions of the $\sys$ system and the real-world web data integration scenarios.
\end{acks}
\bibliography{main}



\end{document}